\documentclass[aps,prd,twocolumn,superscriptaddress]{revtex4-2}

\usepackage{graphicx}
\usepackage{dcolumn}
\usepackage{bm}
\usepackage[colorlinks=true,allcolors=blue]{hyperref}
\usepackage{float}
\usepackage{placeins}
\usepackage{enumerate}
\usepackage{subcaption}
\usepackage{ragged2e}
\usepackage[dvipsnames]{xcolor}
\usepackage{newtxmath}
\usepackage{xspace}
\usepackage{mathtools}
\usepackage{physics}
\usepackage{tensor}
\usepackage[utf8]{inputenc}
\usepackage{acronym}
\usepackage{hyperref}
\usepackage{orcidlink}
\usepackage{color}
\usepackage{units}
\usepackage{todonotes}

\begin{document}

\title{Sub-Solar Mass Intermediate Mass Ratio Inspirals: Waveform Systematics and Detection Prospects with Gravitational Waves}

\author{Devesh Giri \orcidlink{0009-0006-3310-481X}}
\email{devesh.giri@students.iiserpune.ac.in}
\email{deveshgiri.edu@gmail.com}
\affiliation{Indian Institute of Science Education and Research, Pashan, Pune 411008, India}
\affiliation{Nikhef, Science Park 105, 1098 XG, Amsterdam, The Netherlands}

\author{Bhooshan Gadre \orcidlink{0000-0002-1534-9761}}
\email{bhooshan.gadre@iucaa.in}
\affiliation{Inter-University Centre for Astronomy and Astrophysics, Post Bag 4, Ganeshkhind, Pune 411007, India}
\affiliation{Institute for Gravitational and Subatomic Physics (GRASP), Utrecht University, Princetonplein 1, 3584 CC Utrecht, The Netherlands}
\affiliation{Nikhef, Science Park 105, 1098 XG, Amsterdam, The Netherlands}%

\begin{abstract}

We investigate the detectability and waveform systematics of sub-solar mass intermediate mass-ratio inspirals (SSM-IMRIs), characterized by mass ratios $q \sim 10^2-10^4$. Using the black hole perturbation theory surrogate model \textsc{BHPTNRSur1dq1e4} as a reference, we assess the performance of the \textsc{IMRPhenomX} phenomenological family in the high-mass-ratio regime. We find that the inclusion of higher-order gravitational wave modes is critical; their exclusion may degrade the signal-to-noise ratio by factors of $\sim3-5$ relative to quadrupole-only templates. With optimal mode inclusion, SSM-IMRIs are observable out to luminosity distances of $\sim575$ Mpc ($z\sim0.12$) with Advanced LIGO and $\sim10.5$ Gpc ($z\sim1.4$) with the Einstein Telescope. However, we identify substantial systematic uncertainties in current phenomenological approximants. Matches between \textsc{IMRPhenomX} and the reference surrogate model \textsc{BHPTNRSur1dq1e4} degrade to values as low as 0.2 for edge-on inclinations, and fitting factors consistently fall below 0.9, indicating a significant loss of effectualness in template-bank searches. Bayesian parameter estimation reveals that these modeling discrepancies induce systematic biases that exceed statistical errors by multiple standard deviations, underscoring the necessity for waveform models calibrated to perturbation theory in the intermediate mass-ratio regime for robust detection and inference.

\end{abstract}

\maketitle

\acrodef{LSC}[LSC]{LIGO Scientific Collaboration}
\acrodef{LVC}[LVC]{LIGO Scientific and Virgo Collaboration}
\acrodef{LVK}[LVK]{LIGO Scientific, Virgo and KAGRA Collaboration}
\acrodef{aLIGO}{Advanced Laser Interferometer Gravitational-Wave Observatory}
\acrodef{aVirgo}{Advanced Virgo}
\acrodef{LIGO}[LIGO]{Laser Interferometer Gravitational-Wave Observatory}
\acrodef{IFO}[IFO]{interferometer}
\acrodef{BH}[BH]{black hole}
\acrodef{BBH}[BBH]{binary black hole}
\acrodef{BNS}[BNS]{binary neutron star}
\acrodef{IMBH}[IMBH]{intermediate-mass black hole}
\acrodef{NS}[NS]{neutron star}
\acrodef{BHNS}[BHNS]{black hole--neutron star binaries}
\acrodef{NSBH}[NSBH]{neutron star--black hole binary}
\acrodef{PBH}[PBH]{primordial black hole binaries}
\acrodef{CBC}[CBC]{compact binary coalescence}
\acrodefplural{CBC}[CBCs]{compact binary coalescences}
\acrodef{GW}[GW]{gravitational wave}
\acrodef{SSM}[SSM]{subsolar-mass}
\acrodef{IMRI}[IMRI]{intermediate-mass ratio inspiral}
\acrodef{EMRI}[EMRI]{extreme-mass ratio inspiral}
\acrodefplural{EMRIs}[EMRIs]{extreme-mass ratio inspirals}

\acrodef{IMR}[IMR]{inspiral--merger--ringdown}
\acrodef{PSD}[PSD]{power spectral density}
\acrodef{O3}[O3]{third observing run}
\acrodef{O4}[O4]{forth observing run}
\acrodef{CO}[CO]{compact object}
\acrodef{GDB}[GDB]{GraceDB}

\newcommand{\SSM}[0]{\ac{SSM}\xspace}
\newcommand{\IMRI}[0]{\ac{IMRI}\xspace}
\newcommand{\IMRIs}[0]{\acp{IMRIs}\xspace}
\newcommand{\PN}[0]{\ac{PN}\xspace}
\newcommand{\BBH}[0]{\ac{BBH}\xspace}
\newcommand{\BNS}[0]{\ac{BNS}\xspace}
\newcommand{\BH}[0]{\ac{BH}\xspace}
\newcommand{\NR}[0]{\ac{NR}\xspace}
\newcommand{\GW}[0]{\ac{GW}\xspace}
\newcommand{\SNR}[0]{\ac{SNR}\xspace}
\newcommand{\aLIGO}[0]{\ac{aLIGO}\xspace}
\newcommand{\PE}[0]{\ac{PE}\xspace}
\newcommand{\IMR}[0]{\ac{IMR}\xspace}
\newcommand{\PDF}[0]{\ac{PDF}\xspace}
\newcommand{\GR}[0]{\ac{GR}\xspace}
\newcommand{\PSD}[0]{\ac{PSD}\xspace}
\newcommand{\EOS}[0]{\ac{EOS}\xspace}
\newcommand{\LVK}[0]{\ac{LVK}\xspace}
\newcommand{\PYCBC}{\textsc{PyCBC}\xspace}
\newcommand{\BILBY}{\textsc{Bilby}\xspace}
\newcommand{\chie}[0]{\chi_{\rm{eff}}\xspace}
\newcommand{\xas}[0]{\textsc{IMRPhenomXAS}\xspace}
\newcommand{\xhm}[0]{\textsc{IMRPhenomXHM}\xspace}
\newcommand{\bhpt}{\textsc{BHPTNRSur1dq1e4}\xspace}
\newcommand{\Othree}[0]{O3\xspace}
\newcommand{\Ofour}[0]{O4\xspace}
\newcommand{\msun}[0]{\text{M}_{\odot}\xspace}
\newcommand{\ff}{$\mathcal{FF}$\xspace}
\newcommand{\pso}[0]{\texttt{PSO}\xspace}
\newcommand{\diffevol}[0]{\texttt{DiffEvol}\xspace}
\newcommand{\CO}[0]{CO\xspace}
\newcommand{\GDB}[0]{GDB\xspace}

\newcommand{\rough}{\textcolor{red}}
\newcommand{\bug}[1]{\textcolor{blue}{#1}}

\section{\label{sec:intro}INTRODUCTION\protect}

The era of gravitational wave (GW) astronomy has fundamentally transformed our understanding of compact object (CO) populations, with the LIGO-Virgo-KAGRA (LVK) collaboration \cite{LIGO, VIRGO, KAGRA:2020tym} detecting over 200 compact binary coalescences in the fourth observing run (O4) alone \cite{gwtc4_methods, gwtc4_catalog, gwtc4_open_data}. While most of the observed events involve stellar-mass binary black holes \cite{gwtc4_rnp}, a particularly intriguing observational frontier involves sub-solar mass (SSM) compact objects with masses below $1\text{M}_\odot$ \cite{ssm3, ssm1}. Conventional black holes most likely form through the death of massive stars \cite{Oppenheimer:1939ue, Chandrasekhar:1984zz}. These black holes formed in the astrophysical scenario through stellar evolution are not expected in the sub-solar mass range \cite{Chandrasekhar1931}. SSM compact objects cannot arise from known astrophysical processes, making their detection an unambiguous evidence for exotic formation mechanisms.

The most compelling explanation for SSM COs involves primordial black holes (PBHs) which can form in the early universe through gravitational collapse of primordial density \cite{Hawking_collapse, Carr1974, Zel'dovich_primordial, Carr:1975qj, Khlopov:1980mg, dasgupta_pbhs, bird2016, sasaki2016}. These objects could constitute a significant fraction of dark matter and provide unique probes of inflationary physics and the early universe \cite{Baldes:2023rqv, Franciolini2022, Carr:2020gox, Clesse:2016vqa, Clesse:2017bsw, Green:2020jor, Shandera:2018xkn, Afroz:2025urb, Afroz:2024fzp, dasgupta_pbhs, Timmes:1995kp, Sasaki:2018dmp}. There are also studies that suggest a non-primordial formation of SSM COs \cite{Kouvaris:2018wnh} and the existence of neutron stars below the solar mass \cite{Suwa:2018uni}. Enhanced PBH formation during cosmic phase transitions, particularly the QCD epoch, could produce characteristic peaks in the mass spectrum at around $0.01-1\text{M}_\odot$ \cite{Jedamzik2020, byrnes2018}, making SSM searches crucial for understanding both primordial cosmology and the nature of dark matter itself.

Recent analyses of potential SSM candidates have highlighted both the promise and challenges of these searches. The candidate SSM200308, identified in LIGO O3b data with components masses $m_1 = 0.62^{+0.46}_{-0.20} \text{M}_{\odot}$ and $m_2 = 0.27^{+0.12}_{-0.10} \text{M}_{\odot}$, achieved a network signal-to-noise ratio of 8.90 and false-alarm rate of 1 per 5 years\cite{Prunier2023SSM200308}. While not reaching the confidence threshold for definitive detection, this candidate demonstrates the observational accessibility of SSM systems and the need for improved analysis methods to distinguish genuine signals from instrumental noise.

Intermediate mass ratio inspirals (IMRIs) represent a distinct class of GW sources where stellar-mass objects spiral into intermediate-mass black holes (IMBHs) with masses $10^2-10^5 M_{\odot}$, producing mass ratios $q = m_1/m_2$ between $10$ and $10^4$~\cite{Amaro2017,Amaro2018,islam2021,Torres2025}. These systems offer unique advantages for GW astronomy through extended inspiral durations, rich harmonic content, and enhanced parameter estimation precision compared to comparable-mass mergers. The intersection of SSM objects and IMRIs creates sub-solar mass intermediate mass ratio inspirals (SSM-IMRIs), providing the most direct probe of primordial black hole populations while serving as precision laboratories for strong-field gravity tests~\cite{ssm5, Shadykul2023}.

SSM-IMRIs offer several compelling advantages over comparable-mass SSM binaries for GW detection. The intermediate mass ratio extends inspiral duration by factors of $10^2-10^3$, shifting characteristic frequencies into optimal detector sensitivity bands and accumulating larger signal-to-noise ratios over extended observation periods~\cite{ssm5}. Higher-order GW modes become increasingly important for asymmetric systems, with subdominant harmonics contributing up to 50\% of the total detection statistic for extreme mass ratios~\cite{IMRPhenomXHM,Cheung2025IMRI}. Additionally, the strong gravitational fields and diverse orbital dynamics characteristic of IMRIs enable precision tests of general relativity that are impossible with other GW sources.

However, SSM-IMRI searches face formidable challenges that distinguish them from conventional binary searches. Waveform modeling remains problematic, as traditional phenomenological models like \xas and \xhm were calibrated primarily for comparable-mass systems and may introduce systematic biases in the intermediate or extreme mass ratio regime characteristic of SSM-IMRIs~\cite{IMRPhenomXAS,IMRPhenomXHM}. Recent advances in black hole perturbation theory have enabled development of surrogate models such as \bhpt, extending coverage to mass ratios up to $q = 10^4$~\cite{BHPTNRSur1dq1e4,BHPTNRSur2dq1e3}, but significant uncertainties persist regarding systematic errors between different waveform approximations.

Computational challenges present additional barriers to SSM-IMRI detection. Template bank construction for SSM searches requires unprecedented template density, with the recently developed O4 SSM template bank containing over 3.4 million templates covering masses down to $0.2 \text{M}_{\odot}$~\cite{Hanna2024SSMBank}. For a total binary mass of $M$, the computational cost scales approximately as $M^{-8/3}$ for low-mass systems, making traditional matched filtering approaches computationally prohibitive without significant algorithmic innovations. Furthermore, SSM systems exhibit inherently lower GW amplitudes, limiting current Advanced LIGO detection volumes to tens of megaparsecs and requiring careful optimization of search strategies.

Systematic uncertainties between waveform models become particularly problematic for SSM-IMRIs due to the exigent parameter regimes involved. While \bhpt provides theoretical rigor through its foundation in black hole perturbation theory, computational costs limit practical applications to focused parameter estimation rather than broad searches. Phenomenological models offer computational efficiency but may break down in regions of rapid parameter space variation characteristic of high mass-ratio systems \cite{BHPTNRSur1dq1e4,Estelles2021}.

Recent developments have begun addressing these challenges through multiple approaches. The advancement of tidal deformability measurements provides a novel method for distinguishing PBHs from neutron stars or other compact objects in the SSM regime, with current and future detectors capable of measuring tidal effects at the $\mathcal{O}(10\%)$ level for SSM neutron star binaries~\cite{Crescimbeni2024Tidal}. Multiband GW astronomy
combining LISA and ground-based detectors offers transformative prospects for SSM-IMRI science, enabling years-long early inspiral tracking followed by merger detection with unprecedented parameter estimation precision~\cite{ssm5}.

The next-generation detectors will revolutionize SSM-IMRI astronomy through enhanced low-frequency sensitivity and improved strain sensitivity across all frequencies. Einstein Telescope and Cosmic Explorer will extend SSM detection volumes by factors of $10^2-10^3$, transforming these systems from exceptional discoveries to routine observations and enabling population studies essential for constraining primordial black hole abundance~\cite{Maggiore2020,Mukherjee:2021itf}.

The cosmological significance of SSM-IMRI detections extends far beyond confirming exotic object populations. Precise measurements of SSM mass distributions will constrain inflationary models and early universe physics, while merger rate measurements provide direct probes of primordial black hole clustering and formation mechanisms~\cite{Yuan2024PBH,Franciolini2022}. Environmental effects from dark matter spikes and accretion processes may produce detectable signatures in GW signals, offering unique insights into the astrophysical environments surrounding massive black holes \cite{Shadykul2023}.

Current LIGO searches have established meaningful constraints on SSM merger rates, with upper limits ranging from $\sim 10^3$ to $10^6$ Gpc$^{-3}$ yr$^{-1}$ depending on component masses~\cite{ssm3,ssm1}. These constraints are beginning to impact theoretical models of primordial black hole abundance, though broad mass distributions remain consistent with comprising the totality of dark matter in certain scenarios.

The astrophysical implications of SSM-IMRI observations encompass multiple areas of fundamental physics. Tests of general relativity will achieve unprecedented precision through the combination of strong gravitational fields, extended observation duration, and multiple observational bands. Deviations from Einstein's theory may become detectable through systematic analysis of SSM-IMRI populations, particularly in the strong-field regime where alternative gravity theories predict observable modifications~\cite{Berti2015,Yunes2016}.

In this comprehensive study, we address the critical challenges of SSM-IMRI detection and characterization through systematic investigation of waveform systematic uncertainties, optimized search methodologies, and detection prospect assessments. We employ complementary waveform models,  \bhpt for theoretical accuracy and \xhm for computational efficiency, to quantify systematic biases across mass ratios $10^2 < q < 10^3$ with SSM components ranging from $0.2-1.0 \text{M}_{\odot}$. Our analysis incorporates recent advances in higher-order mode modeling, and statistical inference to establish robust frameworks for interpreting future SSM-IMRI observations.

The structure of this paper provides a comprehensive coverage of the science and methodology of SSM-IMRI. Section~\ref{sec:waveforms} reviews a couple of current waveform modeling approaches; namely perturbation theory based surrogate and phenomenological approximations. In Section~\ref{sec:detect}, we study the possibility of detecting SSM-IMRIs with the ground-based GW observatories. Sections~\ref{sec:ff} and~\ref{sec:pe} quantify waveform systematic effects through extensive fitting factor comparisons and parameter estimation studies. We conclude in Section~\ref{sec:conclusions} with synthesis of findings and outlook for SSM-IMRI GW astronomy.

\section{\label{sec:waveforms}Waveforms\protect}

The gravitational waves from a circular binary black hole system observed by a detector are characterized by 15 parameters: 8 intrinsic and 7 extrinsic. The 8 intrinsic parameters are: the two component masses $(m_1, m_2)$ and the two spin vectors $(\mathbf{S_1}, \mathbf{S_2})$. The 7 extrinsic parameters are: inclination angle $\iota$, polarization angle $\psi$, luminosity distance $d_L$, right ascension $\alpha$, declination $\delta$, coalescence time $t_c$ and coalescence phase $\varphi_c$. 

The gravitational wave polarization strain $(h = h_{+} + i h_{\times})$ from a circular non-precessing CBC source is usually expanded into spin-2 weighted spherical harmonic basis (\cite{wave_expand} and Eq. 185. in \cite{Blanchet:2013haa})
\begin{align}
    h(t;\bm{\lambda}, \iota, \varphi_0) = \frac{1}{d_L}\sum_{\ell =2}^{\infty}{\sum_{|m| \leq \ell}}h_{\ell m}(t;\bm{\lambda})Y_{-2}^{\ell m}(\iota,\varphi_0)
\end{align}
where $h_{\ell m}(t;\bm{\lambda})$ are time-dependent multipoles which are functions of intrinsic parameters $\bm{\lambda}$ of the source, and $Y_{-2}^{\ell m}(\iota, \varphi_0)$ are the spin -2 weighted spherical harmonics. $\iota$ denotes the inclination angle of the source which is the angle between the line-of-sight from the detector to the source and the net angular momentum of the binary system. $\varphi_0$ denotes the initial phase of the binary.

The \GW strain observed by the detectors can be written as $h=F_+h_{+}+F_\times h_\times$, where $F_+$ and $F_\times$ denote sky-location and polarisation-dependent detector response functions. The forms of the antenna response functions are dependent on the geometry of the detector and would change depending on the angle subtending by the arms of the interferometric detectors. The antenna response functions for LIGO-Virgo detectors are given by (Eq. 3.29a and 3.29b in \cite{Finn:1992xs}). For detector arms subtending an angle of $\pi/3$ the forms of the antenna pattern functions will be different from that of LIGO (see Eq. 3.6 and 3.7 in \cite{Li:2013lza}).

The relative amplitudes of the different harmonic modes, in general, depend on the inclination $\iota$, polarization $\psi$, masses $(m_1, m_2)$ and spins $(\mathbf{S_1}, \mathbf{S_2})$ of the GW source. The contribution of higher modes relative to the $(2,\pm 2)$ modes generally increases for higher mass ratios and edge-on cases. There is no dependence of the relative amplitudes on the total mass of the system, however, as the noise PSD of the detector depends upon the frequency, a dependence on the total mass can be observed. \cite{measuring_gw_modes}.

Current CBC searches typically use only the dominant $(2, 2)$ mode as the mode contains most of the signal-to-noise (SNR) computational costs. The contribution of higher harmonics becomes increasingly significant with asymmetric mass-ratios \cite{Varma:2014jxa, Kidder:1995zr}. Observation of higher harmonics can break the well-known distance-inclination degeneracy \cite{constrain_incl}. The SSM-IMRIs will have significant higher harmonic content due to the mass ratios ranging from hundreds to thousands.

The accurate modelling of gravitational wave signals from compact binary
mergers requires sophisticated theoretical frameworks that capture the full
inspiral-merger-ringdown (IMR) evolution. For intermediate-mass ratio inspirals
with sub-solar mass companions, the choice of waveform model becomes
particularly critical due to the extreme mass ratios and extended parameter
space coverage required. This study employs two complementary approaches: the
\bhpt model, which leverages black hole perturbation theory for
extreme mass ratios, and the \textsc{IMRPhenomX} family models, which provide
computational efficiency across the broader parameter space.

\subsection{\label{sec:bhpt}\bhpt}

The \bhpt surrogate model~\cite{BHPTNRSur1dq1e4} represents one of the
state-of-the-art for modelling gravitational waveforms from non-spinning binary
black hole systems across comparable to extreme mass ratios, providing coverage
for mass ratios $2.5 \leq q \leq 10^4$. This model is built on a rigorous
foundation in point-particle black hole perturbation theory (ppBHPT), treating
the smaller companion as a point particle orbiting in the Kerr background
spacetime of the primary black hole.

The underlying waveform generation employs a time-domain Teukolsky equation
solver~\cite{Teukolsky1973,Hughes2000}, which computes gravitational
perturbations by solving the linearized Einstein equations in the
Newman-Penrose formalism. The complete inspiral-merger-ringdown evolution is
constructed through three distinct phases: The orbital trajectory is computed
using energy and angular momentum fluxes calculated from gravitational wave
emission, incorporating radiation reaction effects through the balance law
formalism~\cite{Hughes2000}. An updated transition model connects the adiabatic
inspiral to the plunge phase, capturing the dynamics as the system departs from
quasi-circular evolution. The final black hole formation and quasi-normal mode
ringing are modelled using perturbative calculations around the remnant Kerr
spacetime. The model covers waveform durations up to $\sim 30,500 M$ (where $M
= m_1 + m_2$ is the total mass), which may not be sufficient for capturing the
complete observable evolution for SSM-IMRI systems in current detector
bands~\cite{Islam2022}.

\bhpt provides comprehensive spherical harmonic mode content
essential for accurate modelling of asymmetric systems. The uncalibrated
version includes modes up to $\ell = 10$, with over 50 available modes having
$m > 0$, while modes with $m < 0$ are obtained through standard symmetry
relations as the waveform is expressed in terms of spin-weighted spherical
harmonics.

A critical development in the \bhpt model is the incorporation of
calibration to numerical relativity (NR) simulations through an
$\alpha$-$\beta$ scaling technique~\cite{Islam2022,Islam2023cal}. This simple
but effective rescaling procedure corrects for finite-size effects of the
secondary black hole that are neglected in the point-particle approximation
making model ideal for IMRIs: \begin{equation} h^{\ell,m}_{\text{cal}}(t; q) =
\alpha_\ell(q) \, h^{\ell,m}_{\text{ppBHPT}}(\beta(q) t; q), \end{equation}
where $\alpha_\ell(q)$ rescales the mode amplitude and $\beta(q)$ rescales the
time coordinate (and consequently the phase evolution). The $\alpha$-$\beta$
scaling effectively accounts for missing finite-size effects in the
perturbation theory framework~\cite{Islam2023finite}. The amplitude scaling
parameter $\alpha_\ell$ corrects for the difference between point-particle and
extended-body tidal interactions, while the time rescaling $\beta$ adjusts for
the effective gravitational binding energy that differs between the
point-particle and finite-size descriptions.

The calibration parameters are determined by matching ppBHPT waveforms to a
suite of numerical relativity simulations from the SXS and RIT
catalogs~\cite{Boyle2019,Healy2017}. For mass ratios $q = 15$ to $q = 32$, the
calibrated model achieves agreement with numerical relativity at the level of
$\sim 10^{-3}$ in fitting factor for the dominant $(2,2)$ mode. The calibration
is currently implemented for modes up to $\ell = 5$ and mass ratios $q \lesssim
30$, where sufficient numerical relativity waveforms existed then. Beyond these
limits, the model relies on uncalibrated ppBHPT waveforms, introducing
systematic uncertainties that increase with mass ratio and mode order. The
study extending to the equal-mass limit ($q = 1$) demonstrates surprising
agreement between calibrated ppBHPT and NR waveforms even far outside the
expected domain of validity for perturbation theory~\cite{Islam2023cal}.

The surrogate model framework employs reduced-order modelling techniques to
enable rapid waveform evaluation~\cite{Field2014}. Waveform generation times
are typically $10-100$ times larger than phenomenological frequency-domain
models but provide theoretical consistency unavailable in empirical approaches.
The model is publicly available through the \textsc{BHPTNRSurrogate} package as
part of the Black Hole Perturbation Toolkit~\cite{BHPToolkit2024}.

Even though the \bhpt is the-state of the art model for IMRI
applications, it has its limitations. Modes beyond $\ell = 5$ and mass ratios
$q > 30$ rely on uncalibrated ppBHPT, with systematic uncertainties increasing
in these regimes. Also, for mass ratios beyond 30, assume that the extrapolated
calibration is still effective. The $\sim 30,500 M$ waveform duration may be
insufficient for capturing the complete early inspiral for some IMRI systems,
mainly low mass ones like SSM-IMRIs. Even the faster surrogate waveform
generation is orders of magnitude slower than phenomenological models, limiting
applications to focused parameter estimation and extensive template bank
generation for searches. The baseline model applies only to non-spinning
binaries, though recent extensions (BHPTNRSur2dq1e3) incorporate aligned spin
on the primary~\cite{BHPTNRSur2dq1e3}.

Despite these limitations, \bhpt provides the most theoretically
rigorous waveform modelling currently available for intermediate and extreme
mass ratio systems, serving as an essential benchmark for validating
phenomenological models and establishing systematic error budgets for IMRI
science.

\subsection{\label{sec:phenom}\textsc{IMRPhenomX} Family Models}

The IMRPhenomX family represents the current state-of-the-art in
phenomenological waveform modelling, combining computational efficiency with
broad parameter space coverage~\cite{Pratten2020,GarciaQuiros2020}.
For this study, we focus on two specific variants: IMRPhenomXAS for
aligned-spin systems providing only the dominant $(2,\pm 2)$ mode, and
IMRPhenomXHM for higher-order mode inclusion.

IMRPhenomXAS provides the foundational $(2,\pm 2)$ mode modelling with improved
inspiral-merger-ringdown transitions compared to earlier phenomenological
models. The model incorporates effective-one-body (EOB) calibrations and is
tuned to numerical relativity simulations for mass ratios up to $q =
18$~\cite{Pratten2020}. While this mass ratio coverage is significantly
more limited than BHPTNR surrogates, IMRPhenomXAS offers superior computational
efficiency, enabling rapid template generation essential for matched-filtering
searches and parameter estimation studies. The frequency-domain formulation
allows for waveform generation times that are typically 2-3 orders of magnitude
faster than time-domain surrogate models~\cite{Colleoni2023}.

IMRPhenomXHM extends the baseline XAS model to include subdominant harmonics
$(2,1)$, $(3,3)$, $(3,2)$, and $(4,4)$, with mode-mixing effects incorporated
for the $(3,2)$ mode~\cite{GarciaQuiros2020}. The model is calibrated
against hybrid waveforms that combine effective-one-body inspiral evolution
with post-Newtonian amplitudes for subdominant modes, matched to numerical
relativity waveforms and perturbative Teukolsky equation solutions. However,
the calibration for higher-order modes becomes increasingly sparse at high mass
ratios, with reliable coverage typically limited to $q \lesssim 18$ for the
subdominant modes, though some extrapolation to higher mass ratios is
possible~\cite{Estelles2022,Estelles2021}.

For asymmetric mass ratio systems characteristic of SSM-IMRIs, IMRPhenomXHM
faces particular challenges in accurately modelling the complex interplay
between higher-order modes and extreme mass ratios. The phenomenological
framework relies on smooth interpolation across parameter space, which may
break down in regions of rapid variation characteristic of high-$q$ systems.
The calibration to numerical relativity becomes increasingly sparse at high
mass ratios, potentially introducing systematic biases in parameter recovery.

\subsection{Waveform model selection}

The choice of the BHPTNR surrogate and IMRPhenomX models for this study represents
an optimal balance between accuracy and computational feasibility for SSM-IMRI
characterization. BHPTNR surrogates provide the theoretical rigour and extended
mass ratio coverage essential for accurate modelling of true IMRI systems, with
their foundation in first-principles black hole perturbation theory ensuring
physical consistency across the extreme parameter space. The comprehensive mode
content up to $\ell = 10$ (uncalibrated) and $\ell = 5$ (calibrated) enables
accurate representation of the complex gravitational wave emission patterns
characteristic of highly asymmetric systems~\cite{Islam2022}.

Meanwhile, IMRPhenomX models enable comprehensive parameter space exploration
and statistical analysis through their exceptional computational efficiency.
The rapid waveform generation capabilities make them practical for extensive
Monte Carlo simulations, template bank construction, and Bayesian parameter
estimation campaigns that would be computationally prohibitive with BHPTNR
surrogates. Despite their limitations at extreme mass ratios, the
phenomenological models provide a crucial computational bridge that enables
systematic studies of detectability and parameter estimation accuracy across
the full SSM-IMRI parameter space~\cite{Pratten2020,GarciaQuiros2020}.

This dual-model approach allows for critical comparison of systematic
uncertainties and validation of phenomenological model reliability in the
extreme parameter regime. By quantifying fitting factors and parameter
estimation biases between these fundamentally different modelling
approaches—one based on rigorous perturbation theory and the other on
phenomenological interpolation—we can establish confidence bounds on
astrophysical inference from future SSM-IMRI detections. The computational
efficiency of IMRPhenomX models enables extensive population synthesis studies,
while the theoretical accuracy of BHPTNR surrogates provides the benchmark for
systematic error quantification.

Furthermore, the complementary nature of these models' systematic uncertainties
provides a robust framework for error estimation. Where BHPTNR models may
suffer from calibration uncertainties at comparable mass ratios and
computational limitations, IMRPhenomX models provide rapid evaluation but with
reduced accuracy at extreme mass ratios. By utilising both approaches and
quantifying their agreement across the parameter space, we can identify regions
where systematic uncertainties are minimised and establish reliable bounds on
the precision of SSM-IMRI characterisation with current gravitational wave
detection methods.

\begin{table*}
    \centering
    \caption{Allowed values for some of the parameters of the models. None of the models includes spin-precession and includes only aligned spin (if any).}
    \label{tab:wave_models}
    \begin{tabular}{c|c|c|c|c}
        \hline
        Model & Spin & $q_\text{max}$ & $\ell_\text{max}$ & Modes $(\ell, |m|)$\\
        \hline
        \hline
        \textsc{IMRPhenomXAS} & \checkmark & $1000$ & $2$ & (2, 2) \\
        \hline
        \textsc{IMRPhenomXHM} & \checkmark & $1000$ & $4$ & (2,\{2, 1\}), (3, \{3, 2\}), (4, 4) \\
        \hline
        \textsc{BHPTNRSur1dq1e4} \small{(calibrated)}& $\times$ & $10000$ & $5$ & (2, \{2, 1\}), (3, \{3, 2, 1\})\\
        & & & & (4, \{4, 3, 2\}), (5, 5) \\
        \hline
        \textsc{BHPTNRSur1dq1e4} \small{(uncalibrated)}& $\times$ & $10000$ & $10$ & (2, \{2, 1\}), (3, \{3, 2, 1\})\\
        & & & & (4, \{4, 3, 2\}), (5, \{5, 4, 3\}) \\
        & & & & (6, \{6, 5, 4\}), (7, \{7, 6, 5\}) \\
        & & & & (8, \{8, 7, 6\}), (9, \{9, 8, 7\}) \\
        & & & & (10, \{10, 9\}) \\
        \hline
    \end{tabular}
\end{table*}

Table (\ref{tab:wave_models}) tabulates some of the allowed parameters for the models used in our analyses. Note that for a given $\ell_\text{max}$ value, the models do not include all $|m| \leq \ell \leq \ell_\text{max}$. \textsc{IMRPhenomXAS} and \textsc{IMRPhenomXHM} are frequency-domain waveforms, while the \textsc{BHPTNRSur1dq1e4} is a time-domain waveform. In our analyses, we use either only the $l=2$, $|m|=2$ mode or all the modes generated by the models. For all the waveforms generated using \textsc{BHPTNRSur1dq1e4}, $\ell_\text{max}=5$ for NR-calibrated case, and $\ell_\text{max}=10$ for uncalibrated case, unless otherwise stated.

\section{\label{sec:detect}Detectability\protect}

\begin{figure*}[ht]
\begin{subfigure}{0.45\textwidth}
  \centering
  \includegraphics[width=\linewidth]{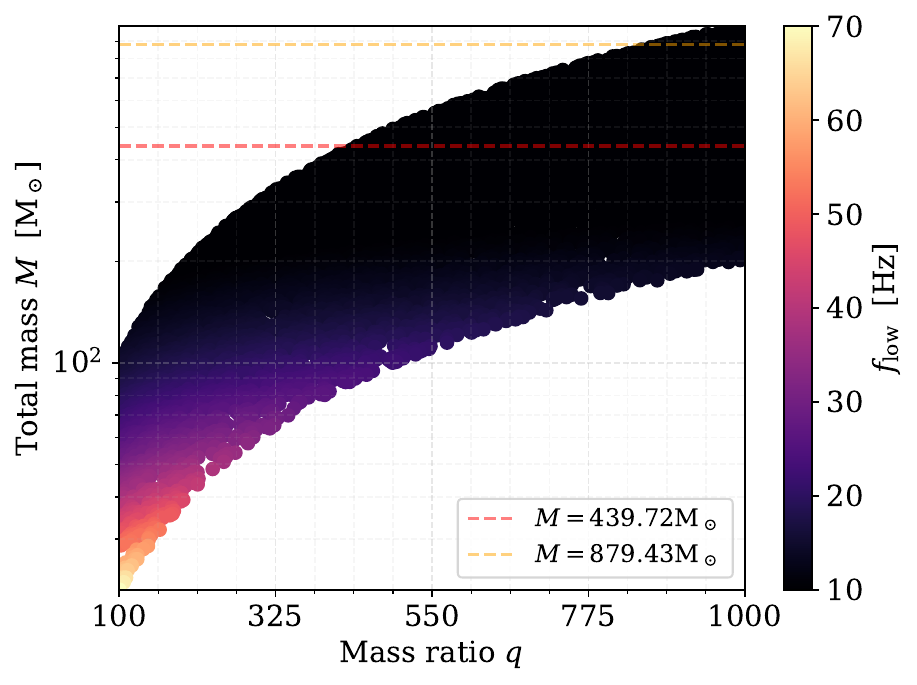}  
  \caption{all calibrated modes}
  \label{fig:f_low_aLIGO_all_calibrated_modes}
\end{subfigure}
\begin{subfigure}{0.45\textwidth}
  \centering
  \includegraphics[width=\linewidth]{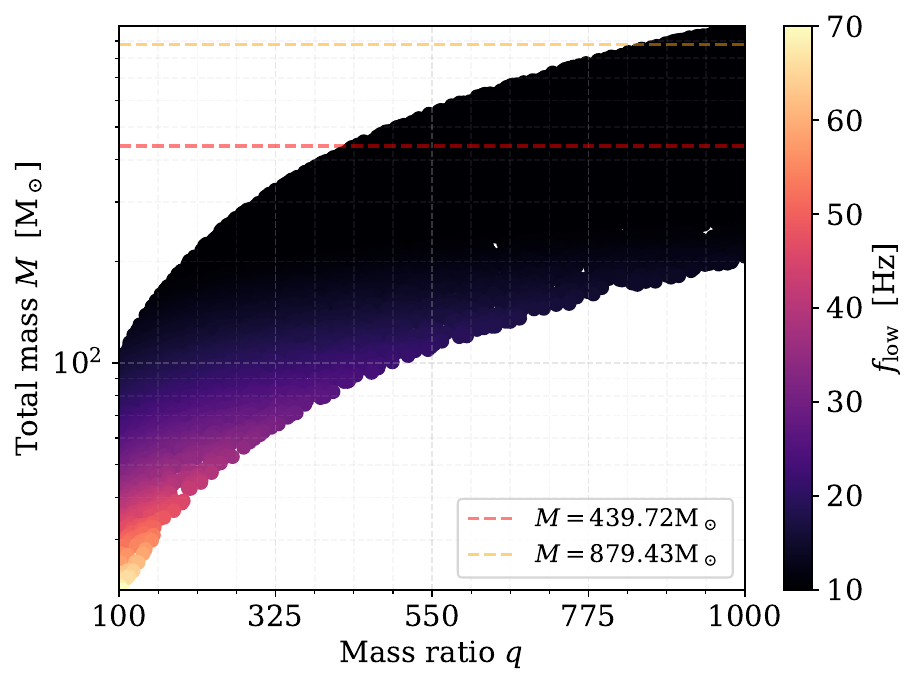}  
  \caption{only $(2,\pm 2)$ calibrated modes}
  \label{fig:f_low_22_calibrated_modes}
\end{subfigure}
\caption{\justifying{Low-frequency cutoff, $f_\text{low}$, for different cases in the $q\text{-}M$ plane for computations done for the case of aLIGO sensitivity. The maximum between $10$Hz and the starting frequency of the \bhpt waveform (for both all calibrated modes and only $(2,\pm 2)$ calibrated modes) is chosen as $f_\text{low}$. $f_\text{ISCO}$ is $10$Hz for $M=439.72\text{M}_\odot$ and $5$Hz for $M=879.43\text{M}_\odot$}.}
\label{fig:f_low_plot}
\end{figure*}

\begin{figure*}[ht]
\begin{subfigure}{0.45\textwidth}
  \centering
  \includegraphics[width=\linewidth]{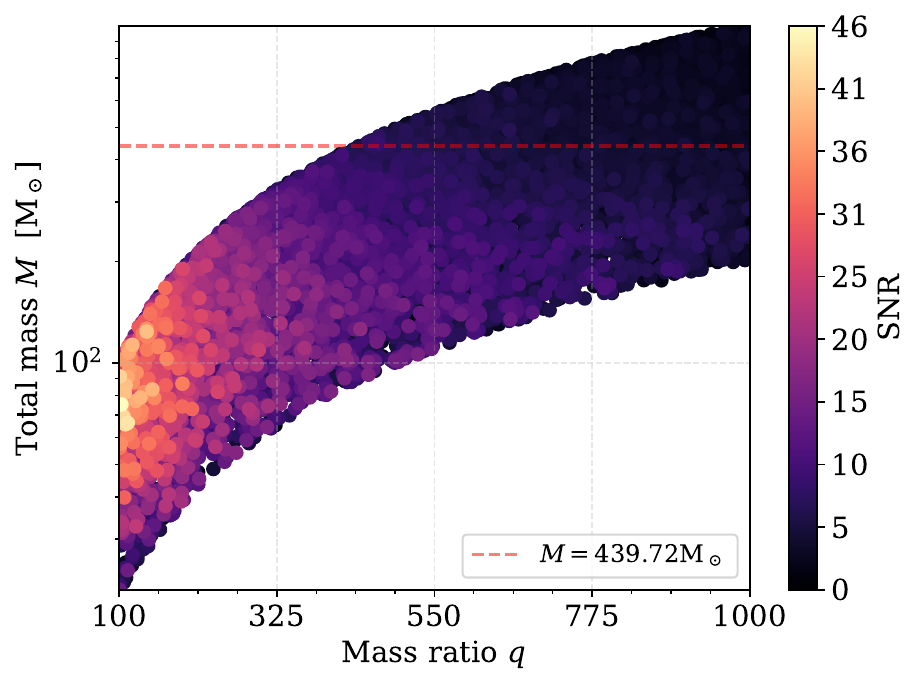}  
  \caption{aLIGO}
  \label{fig:snrSur_q_M_aLIGO_all_calib_modes}
\end{subfigure}
\begin{subfigure}{0.45\textwidth}
  \centering
  \includegraphics[width=\linewidth]{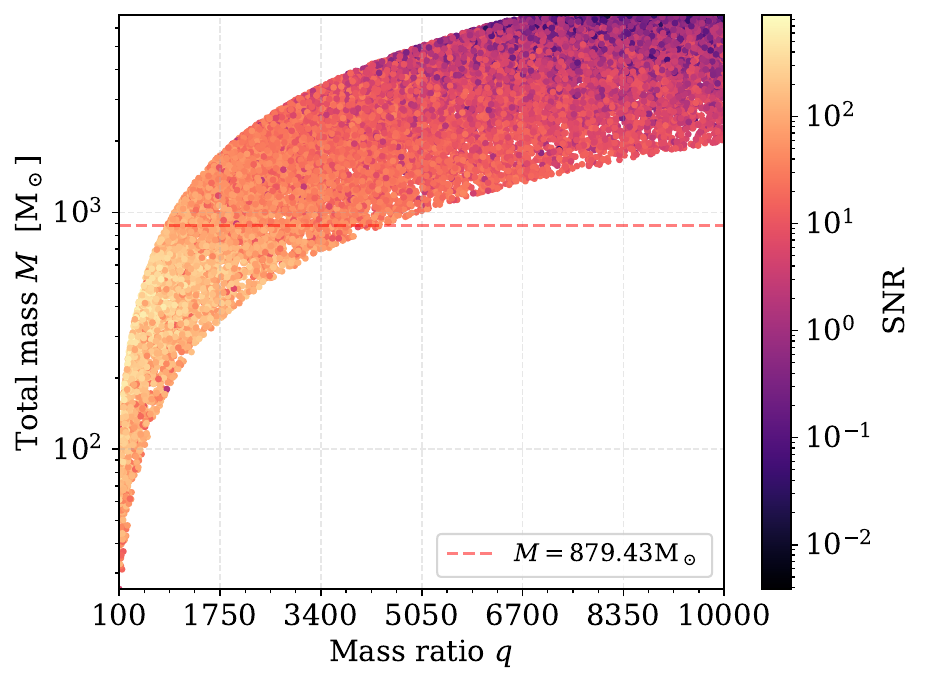}    
  \caption{ET}
  \label{fig:snrSur_q_M_ET_all_calib_modes}
\end{subfigure}
\caption{\justifying{SNR as color-code in the $q-M$ plane calculated using \bhpt (all calibrated modes, i.e., till $l_\text{max}=5$) for aLIGO and ET sensitivities. \ref{fig:snrSur_q_M_aLIGO_all_calib_modes} and \ref{fig:snrSur_q_M_ET_all_calib_modes} show the SNR for aLIGO and ET, respectively. $d_L=100$ Mpc for all the cases.}}
\label{fig:aLIGO_snr_plot}
\end{figure*}

The observability of a GW signal is dictated by its loudness and the sensitivity of the detectors; quantified together as signal-to-noise ratio (SNR). The optimal SNR $\rho_\text{opt}(\bm{\theta})$ of a GW source characterized by parameters $\bm{\theta}$ is given by
\begin{align}
    \rho_\text{opt}(\boldsymbol{\theta}) = {\langle h,h \rangle \label{snr_eq}}^{1/2}
\end{align}
where $\langle\cdot,\cdot\rangle$ denotes the noise-weighted scalar product with a noise power spectral density (PSD), $S_n(f)$, and is given as
\begin{align}
    \langle a,b \rangle = 4 \Re \int_{f_{\text{min}}}^{f_{\text{max}}}{\frac{\Tilde{a}^*(f;\boldsymbol{\theta})\Tilde{b}(f;\boldsymbol{\theta})}{S_n(f)}df} \label{scalar_prod}.
\end{align}
$\sim$ at the top denotes the Fourier transform of the time series signal and $*$ denotes complex conjugation.

To study detectability of SSM-IMRIs, we simulate systems by uniformly sampling the mass ratio $q$, the secondary mass $m_2$, inclination $\iota$, polarization $\psi$, right ascension $\alpha$ and declination $\delta$ in the ranges: $100 \leq q \leq 1000$ (for aLIGO) and $100 \leq q \leq 10000$ (for ET), $0.2\text{M}_\odot \leq m_2 \leq 1\text{M}_\odot$, $0 \leq \iota \leq \pi$, $0 \leq \psi \leq 2\pi$, $0 \leq \alpha \leq 2\pi$, and $-\pi/2 \leq \delta \leq \pi/2$. We consider only non-spinning systems for the current study. For each sample, we generate waveforms using \bhpt, \xhm and \xas models (described in Section \ref{sec:waveforms}) and project them onto the LIGO Hanford and the Einstein Telescope using their respective antenna pattern functions to get a strain time series as observed by the respective detectors. For each of the cases, we compute the optimal SNR with the projected strain and the noise PSD for each detector.\footnote{The noise PSDs used for SNR calculations are \texttt{aLIGOZeroDetHighPower}, and \texttt{EinsteinTelescopeP1600143} for LIGO-Hanford and ET respectively.} We have considered IMRI signals with duration less than $1000$s for aLIGO and less than $2000$s for ET as the main motivation for this work is to assess the possibility of the detection of SSM-IMRIs in the current and next generational GW detectors.  We use \bhpt in 4 configurations: 
(i) only $(2,\pm 2)$ modes and calibrated, 
(ii) all calibrated modes, 
(iii) only $(2,\pm 2)$ modes and uncalibrated, and 
(iv) all uncalibrated modes. 

The duration of the \bhpt model is fixed at $30500m_1$ in the geometric units, where $m_1$ is the mass of the primary. Since the length of the waveform is fixed, the lowest possible reference frequency of the $(2, \pm2)$ mode changes with the primary mass. We set the low-frequency cutoff for SNR computation for \bhpt as the maximum between $10$ Hz ($5$ Hz) and the starting frequency of the \bhpt model for aLIGO (ET) for each case.

Fig. \ref{fig:f_low_plot} shows the variation of the low frequency cutoff, $f_\text{low}$, in the $q\text{-}M$ plane. Fig. \ref{fig:f_low_aLIGO_all_calibrated_modes} shows the variation for the cases in which all the calibrated modes of \bhpt have been considered, and Fig. \ref{fig:f_low_22_calibrated_modes} in which only the $(2,\pm 2)$ modes have been considered. The frequency, $f_{\text{ISCO}}$, at the Innermost Stable Circular Orbit (ISCO), is inversely proportional to the total mass of the binary. For $M>439.72\text{M}_\odot$, $f_\text{ISCO} < 10$Hz and for $M>879.43\text{M}_\odot$, $f_\text{ISCO} < 5$Hz. Thus, we can not use only the $(2,\pm2)$ modes for SSM-IMRI searches with high total masses. For asymmetric mass ratio cases, the next-to-leading mode can contribute within an order of magnitude of the quadrupole mode, as shown in \cite{shoemaker_higher_modes} using numerical relativity waveforms. \cite{Varma:2014jxa} discusses the signals with $20\text{M}_\odot \leq M \leq 250\text{M}_\odot$ and $1\leq q \leq 18$, and shows that neglecting higher harmonics will introduce systematic error larger than $1\sigma$ statistical errors for $q\gtrsim 4$ and $M\gtrsim 150\text{M}_\odot$. Neglecting higher harmonics can lead to losses of more than $10\%$ of events in the case of aLIGO for $q\geq 6$ and $M\geq 100\text{M}_\odot$, as discussed in \cite{CalderonBustillo:2015lrt}. While previous studies have highlighted the importance of higher harmonics in the cases of asymmetric binary black hole systems, we study the effects of higher harmonics in very asymmetric black hole binaries ($q\gtrsim 100$) with a sub-solar mass component.

Fig. \ref{fig:aLIGO_snr_plot} shows the SNR with \bhpt using all the calibrated modes for different mass-ratio and total mass cases for advanced LIGO (\ref{fig:snrSur_q_M_aLIGO_all_calib_modes}) and ET (\ref{fig:snrSur_q_M_ET_all_calib_modes}) sensitivities and luminosity distance $d_L = 100$ Mpc. Fig. \ref{fig:snrSur_q_M_aLIGO_all_calib_modes} shows that the SNR decreases with increasing total mass and mass ratio. The loud IMRIs have lower mass-ratio cases and total mass in $50\text{M}_\odot<M<150\text{M}_\odot$ for which $f_\text{ISCO}$ is in the range $30-100$ Hz. More than $99\%$ of the cases up to the total mass $M<300\text{M}_\odot$ have $\text{SNR}>8$. More than $99\%$ of the cases up to the mass ratio $q<700$ have $\text{SNR}>8$. As shown in Fig. \ref{fig:snrSur_q_M_ET_all_calib_modes}, at $100$ Mpc, ET covers a much larger parameter space and pushes up to $q=10000$ with detectable SNRs. We can consider even higher and more asymmetric cases for ET sensitivity. This is interesting because we can study heavier IMBHs, which are in the frequency band of the ground-based detectors and inaccessible to LISA. In both panels of Fig. \ref{fig:aLIGO_snr_plot}, the trend is not absolutely smooth and has some low SNR cases mixed with the cases of high SNR region because we have not averaged over the inclination angle. Later, we show that the inclination $\iota$ plays an important role and significantly affects the SNR for IMRIs. The total mass corresponding to the loud SNR shifts to higher values in ET as compared to aLIGO, as ET is sensitive to the lower frequencies than aLIGO. 
Since the maximum \bhpt duration is fixed at $30000 m_1$; we only get the partial waveform as we lose the low-frequency content. This includes the loss in some of the higher modes. So, we discuss the effects of higher harmonics in detail further.

\begin{figure*}[ht]
\begin{subfigure}{0.45\textwidth}
  \centering
  \includegraphics[width=\linewidth]{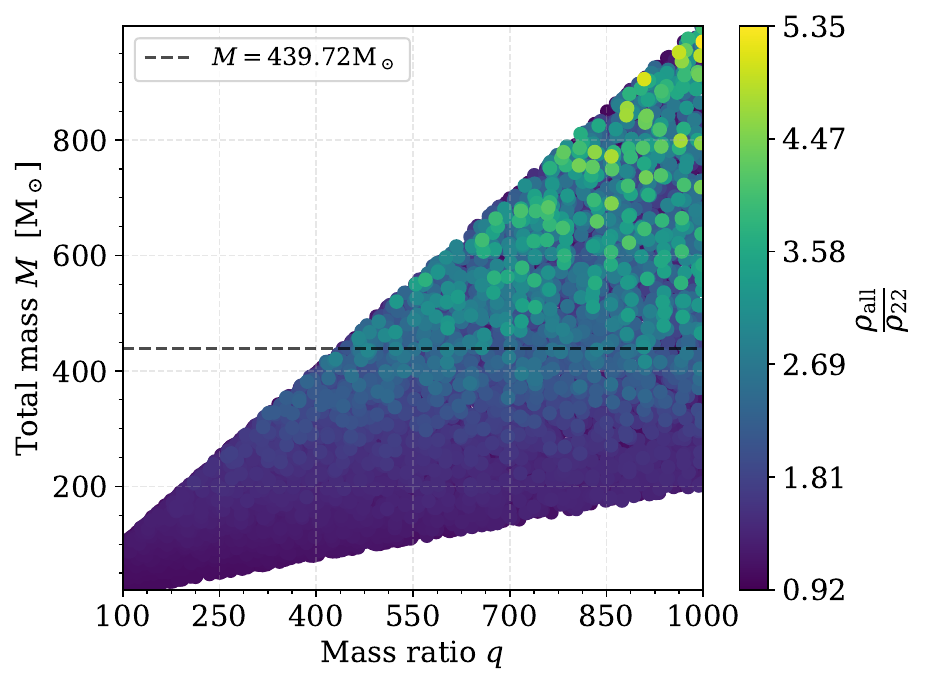}  
  \caption{aLIGO}
  \label{fig:higher_mode_q_M_aLIGO_all_calib}
\end{subfigure}
\begin{subfigure}{0.45\textwidth}
  \centering
  \includegraphics[width=\linewidth]{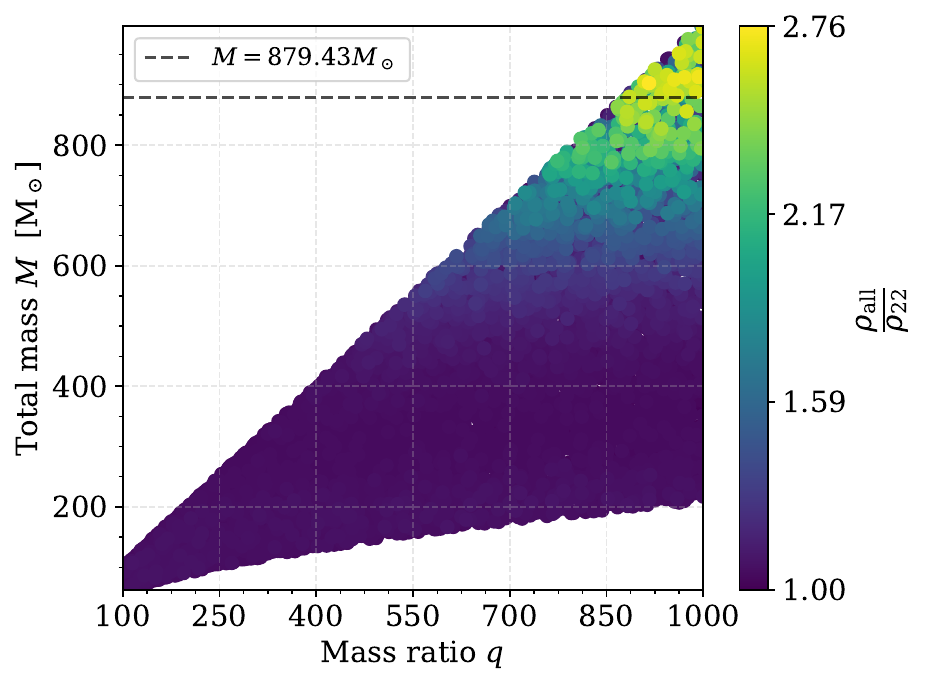}    
  \caption{ET}
  \label{fig:higher_mode_q_M_ET_all_calib}
\end{subfigure}
\caption{\justifying{Ratio of SNRs using all the calibrated modes and only the $(2,\pm2)$ modes of \bhpt shown in the $q-M$ plane. \ref{fig:higher_mode_q_M_aLIGO_all_calib} and \ref{fig:higher_mode_q_M_ET_all_calib} show the ratio of SNRs for aLIGO and ET, respectively, for the same parameter space.}}
\label{fig:higher_mode_qm_plot}
\end{figure*}

\begin{figure*}[ht]
\begin{subfigure}{0.45\textwidth}
  \centering
  \includegraphics[width=\linewidth]{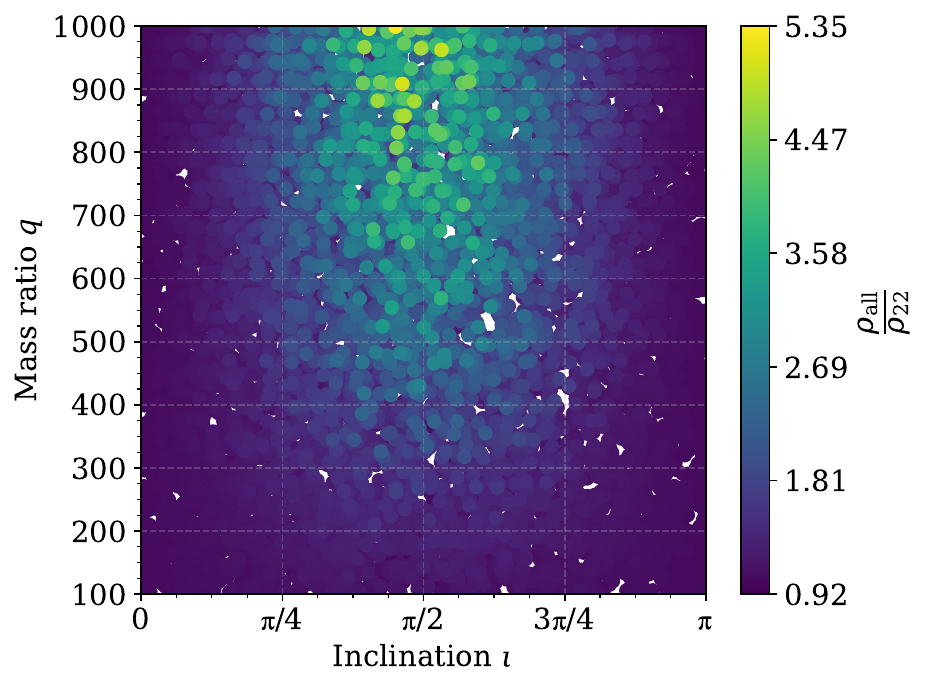}  
  \caption{aLIGO}
  \label{fig:higher_mode_inc_q_aLIGO_all_calib}
\end{subfigure}
\begin{subfigure}{0.45\textwidth}
  \centering
  \includegraphics[width=\linewidth]{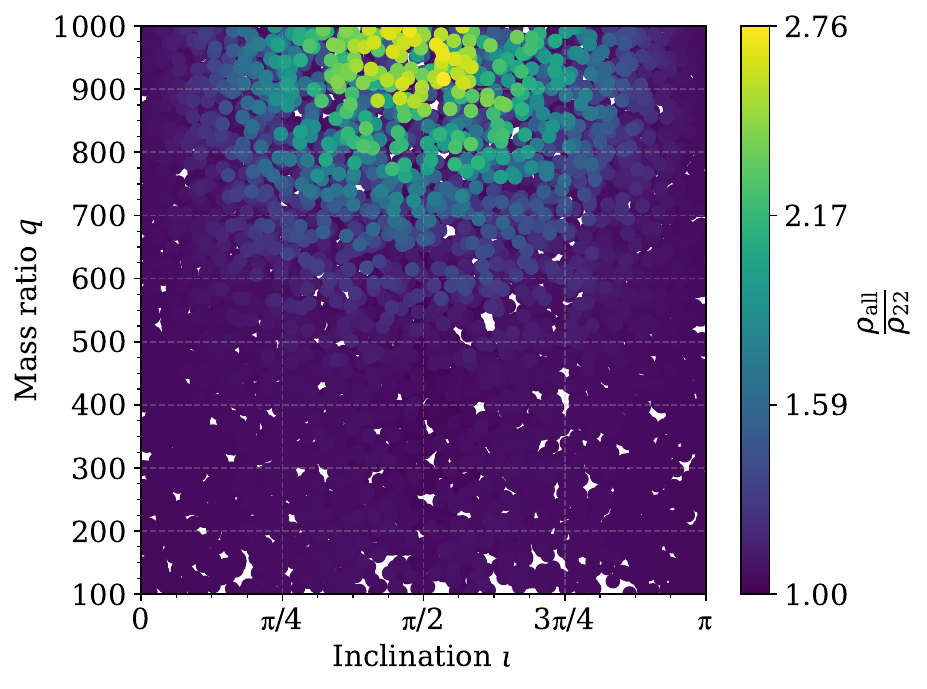}    
  \caption{ET}
  \label{fig:higher_mode_inc_q_ET_all_calib}
\end{subfigure}
\caption{\justifying{Dependence of the higher modes on the inclination angle. Each case is same as that in Fig. \ref{fig:higher_mode_qm_plot}}}
\label{fig:higher_mode_iq_plot}
\end{figure*}

Due to asymmetric masses, GW signals from IMRIs are rich in harmonic content, and their relative ratios depend strongly on the inclination angle. Fig. \ref{fig:higher_mode_qm_plot} shows the ratio of SNRs from all the calibrated modes to only the $(2,\pm2)$ modes of \bhpt in the $q-M$ plane. The contribution to the overall SNR from higher modes increases with increasing mass and mass ratio, as expected, and can be as high as 400\% for aLIGO. The higher mode content becomes significant for the total mass greater than 300, as can be seen in Fig.~\ref {fig:higher_mode_q_M_aLIGO_all_calib}. For $M>439.72\text{M}_\odot$, $f_\text{ISCO}$ drops below the low frequency cutoff of $10Hz$ for aLIGO and $(2,\pm2)$ mode content diminishes. However, higher harmonics make a sufficient contribution to this region of the parameter space and can be used to study such systems. Similarly, for ET,  Fig. \ref{fig:higher_mode_q_M_ET_all_calib} shows the ratio of SNRs for the same parameter space. As ET has a higher sensitivity at lower frequencies, higher mode contributions are less significant for the parameter space being considered as compared to aLIGO. For aLIGO, we also find that SNR using all calibrated modes can be less than the SNR using only the $(2,\pm2)$ modes of \bhpt for a few SSM-IMRI cases. These are the cases of high mass-ratio systems with face-on inclinations (i.e. $\iota$ is close to $0$ or $\pi$).

Fig. \ref{fig:higher_mode_iq_plot} shows the ratio of SNRs as discussed above in the mass ratio-inclination ($q-\iota$) plane for aLIGO and ET. The plots show the dependence of higher mode content on the inclination angle. Higher modes contribute significantly to edge-on cases and have minimal contribution to face-on cases as expected.

However, the maximum SNR comes from the face-on cases, not the edge-on ones, at the fixed luminosity distance, which makes face-on cases detectable at larger distances. Now, we study systematics between \bhpt and \textsc{IMRPhenomX} waveform models.

\begin{figure}[ht]
    \centering
    \includegraphics[width=\linewidth]{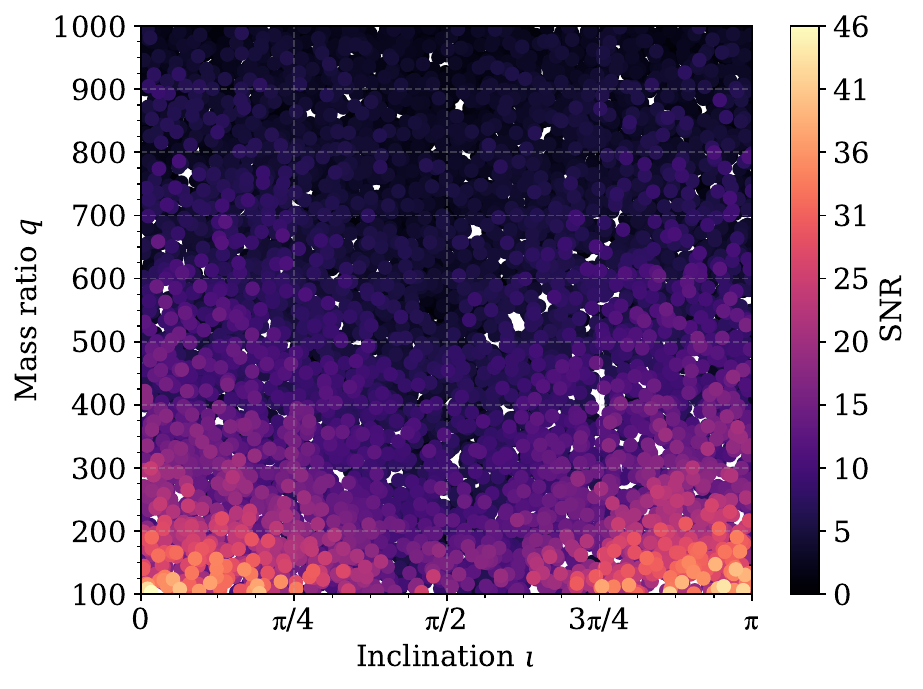}
    \caption{\justifying{Dependence of SNR on the inclination angle $\iota$. SNR is calculated using all the calibrated modes of \bhpt for aLIGO sensitivity.}}
    \label{fig:snr_inc_aLIGO_all_calib}
\end{figure}
Fig. \ref{fig:snr_inc_aLIGO_all_calib} shows the dependence of the higher mode contribution to the SNR as a fraction of $(2, \pm 2)$ mode SNR on the inclination angle computed using all the calibrated modes of \bhpt. For $q > 300$, higher modes contribute significantly to the overall SNR when IMRIs are not face-on. 

\begin{figure*}[ht]
\begin{subfigure}{0.31\textwidth}
  \centering
  \includegraphics[width=\linewidth]{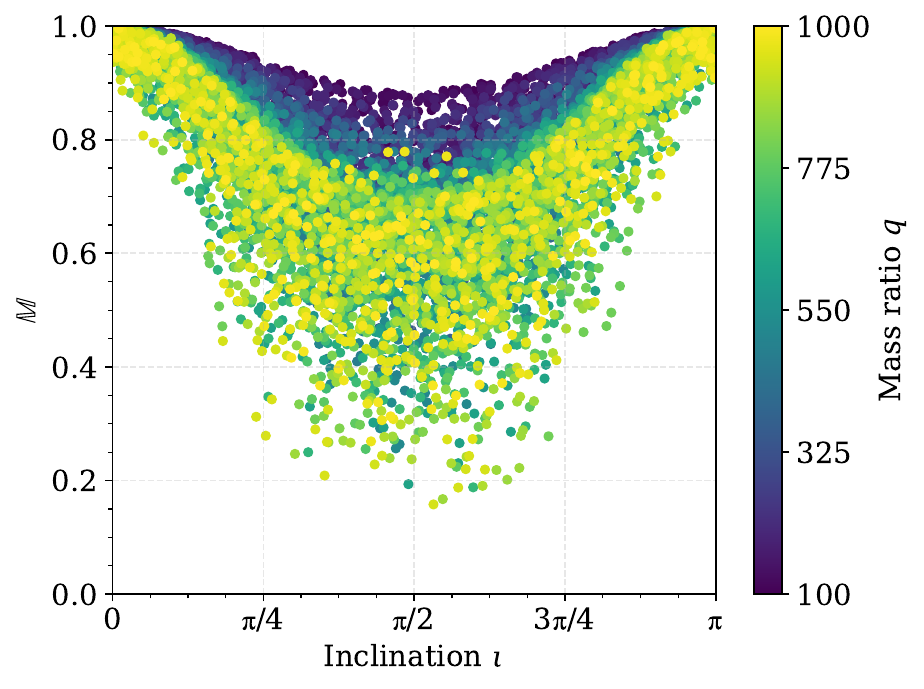}  
  \caption{XAS-XHM}
  \label{fig:xas_xhm_h1_all_calib}
\end{subfigure}
\begin{subfigure}{0.31\textwidth}
  \centering
  \includegraphics[width=\linewidth]{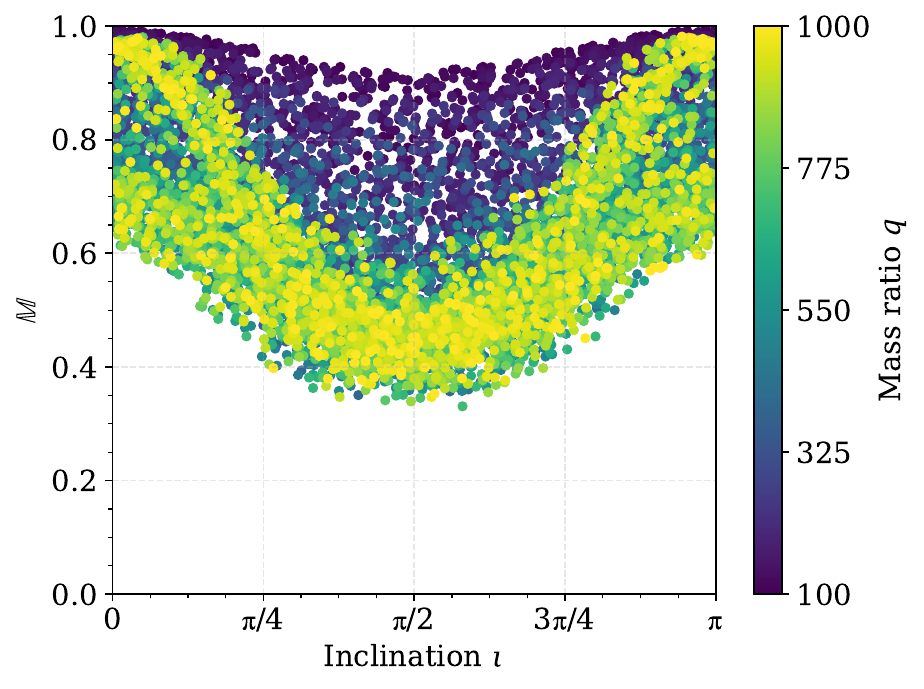}    
  \caption{XAS-\bhpt}
  \label{fig:xas_sur_h1_all_calib}
\end{subfigure}
\begin{subfigure}{0.31\textwidth}
  \centering
  \includegraphics[width=\linewidth]{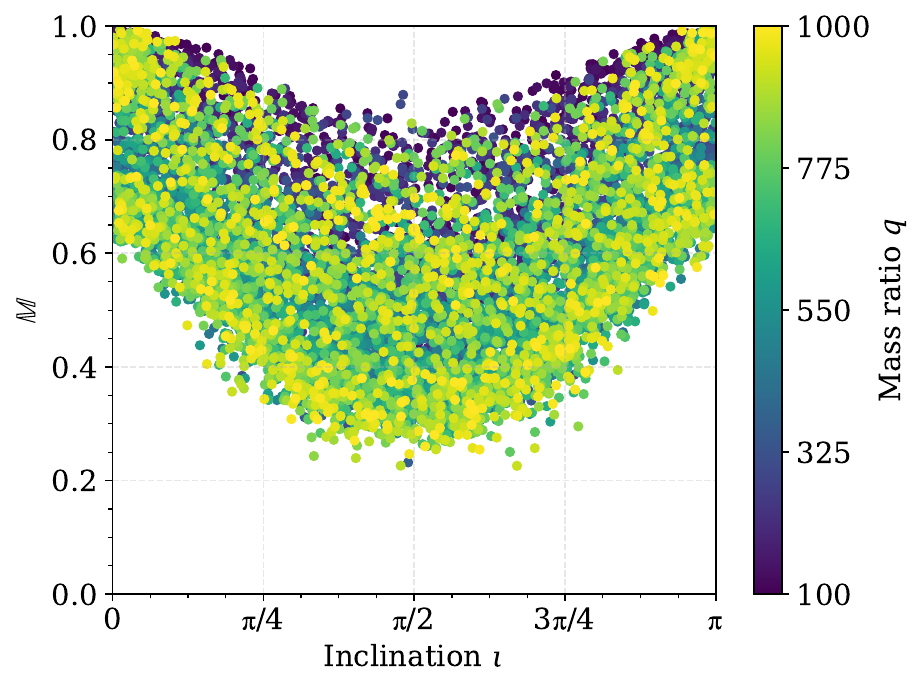}    
  \caption{XHM-\bhpt}
  \label{fig:xhm_sur_h1_all_calib}
\end{subfigure}
\begin{subfigure}{0.31\textwidth}
  \centering
  \includegraphics[width=\linewidth]{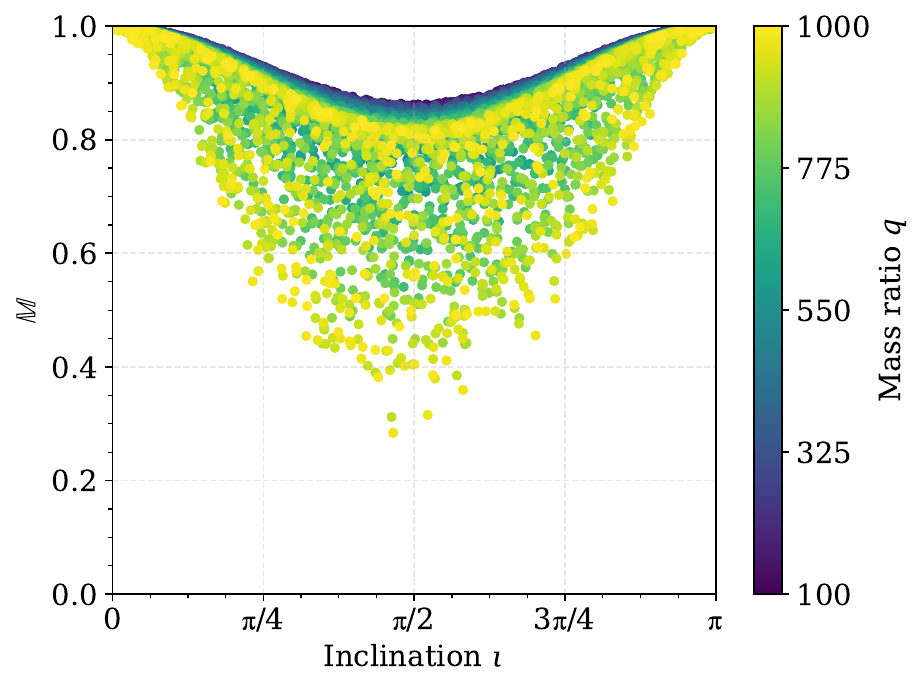}  
  \caption{XAS-XHM}
  \label{fig:xas_xhm_e1_all}
\end{subfigure}
\begin{subfigure}{0.31\textwidth}
  \centering
  \includegraphics[width=\linewidth]{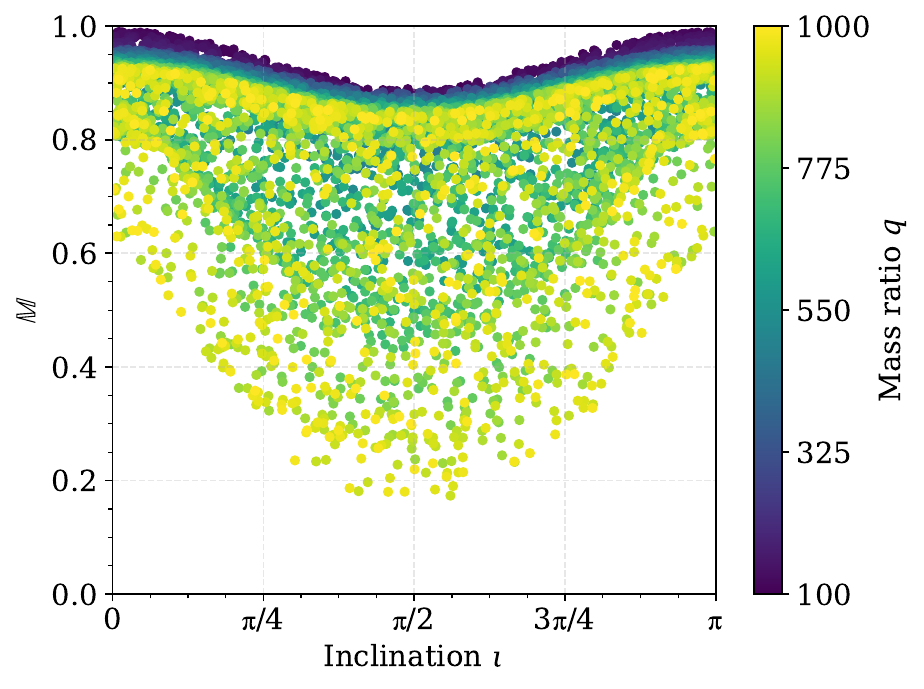}    
  \caption{XAS-\bhpt}
  \label{fig:xas_sur_e1_all}
\end{subfigure}
\begin{subfigure}{0.31\textwidth}
  \centering
  \includegraphics[width=\linewidth]{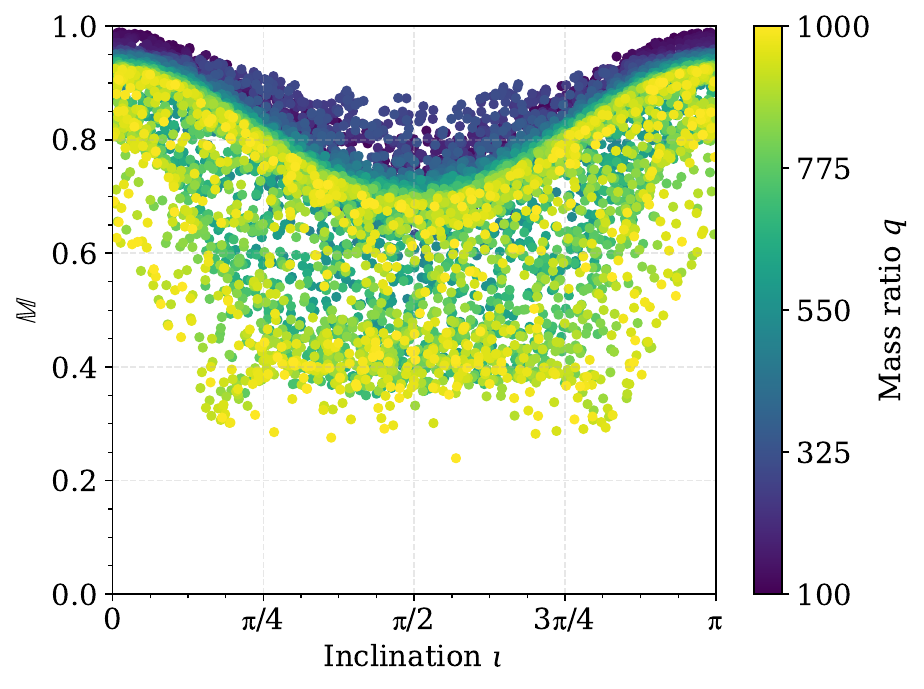}    
  \caption{XHM-\bhpt}
  \label{fig:xhm_sur_e1_all}
\end{subfigure}
\caption{\justifying{Match between the waveforms \xas, \xhm and \bhpt assuming aLIGO sensitivity (for \ref{fig:xas_xhm_h1_all_calib}, \ref{fig:xas_sur_h1_all_calib} and \ref{fig:xhm_sur_h1_all_calib}) and ET sensitivity (for \ref{fig:xas_xhm_e1_all}, \ref{fig:xas_sur_e1_all} and \ref{fig:xhm_sur_e1_all}). The colorbar represents the mass ratio $q$ varying from $100$ to $1000$. All calibrated modes ($\ell_\text{max}=5$) have been used for \bhpt waveform models for match computations. The low frequency cutoff follows Fig. \ref{fig:f_low_plot}.}}
\label{fig:match_bw_waveforms}
\end{figure*}

\section{Waveform systematics}

In this section, we compare \textsc{IMRPhenomX} waveforms against calibrated and uncalibrated \bhpt waveforms to understand and quantify systematic differences in the SSM-IMRI parameter space. As the motivation of the study is the detection of such IMRIs, we will employ the match and fitting factor as figures of merit apart from parameter estimation. We perform the computations for $q < 1000$ due to the unavailability of \xas and \xhm models for $q>1000$.

\subsection{Match}
The match ($\mathbb{M}$) quantifies how well two given waveform models agree at the given value of parameters used for simulation.
The match is defined as the overlap, $\mathcal{O}$, maximised over the coalescence phase and time, for any two arbitrary time series $a$ and $b$. This can be written as
\begin{align}
    \mathbb{M}(a,b) = \max_{t_c, \varphi_c}\mathcal{O}(a,b) \label{match_def}
\end{align} where
\begin{align}
    \mathcal{O}(a,b) = \langle \hat{a},\hat{b} \rangle = \frac{\langle a,b \rangle}{\langle a,a \rangle \langle b,b \rangle}. \label{overlap_eq}
\end{align}

Again, we sample IMRI parameters uniformly as described in Sec .~\ref{sec:detect} with restricted mass ratios.
Fig. \ref{fig:match_bw_waveforms} shows match values between different pairs of waveform models as a function of inclination and mass ratio. The top row corresponds to aLIGO and the bottom one is for ET. The columns correspond to matches between the two \textsc{IMRPhenomX} models (\xas and \xhm), \bhpt against the phenomenological quadrupolar model (\xas) and \bhpt against the phenomenological model that includes higher harmonics (\xhm). All the plots show the increasing mismatches with more asymmetrical IMRIs, and for edge-on configurations, as this is where higher mode contributions becomes most significant. Again, \xas against \xhm performs better in ET than in aLIGO, as better lower frequency sensitivity diminishes the impact of higher modes for the total masses we are considering. For aLIGO: Fig. \ref{fig:xas_xhm_h1_all_calib} shows the match between \xas and \xhm in which about $68\%$ of the cases have the match less than $0.9$ and about $70\%$ of such cases lie in $\pi/4 < \iota < 3\pi/4$ (edge-on region); Fig. \ref{fig:xas_sur_h1_all_calib} shows the match between \xas and \bhpt in which about $90\%$ of the cases have the match less than $0.9$ and about $54\%$ of such cases lie in the edge-on region; Fig. \ref{fig:xhm_sur_h1_all_calib} shows the match between \xhm and \bhpt in which about $93\%$ of the cases have the match less than $0.9$ and about $53\%$ of such cases lie in the edge-on region. Similarly, for ET: Fig. \ref{fig:xas_xhm_e1_all} shows the match between \xas and \xhm in which about $58\%$ of the cases have the match less than $0.9$ and about $79\%$ of such cases lie in $\pi/4 < \iota < 3\pi/4$ (edge-on region); Fig. \ref{fig:xas_sur_e1_all} shows the match between \xas and \bhpt in which about $74\%$ of the cases have the match less than $0.9$ and about $65\%$ of such cases lie in the edge-on region; Fig. \ref{fig:xhm_sur_e1_all} shows the match between \xhm and \bhpt in which about $85\%$ of the cases have the match less than $0.9$ and about $59\%$ of such cases lie in the edge-on region. We have used the $(2,2)$ mode match, and the maximization over the phase $\varphi_c$ for waveform models with higher harmonics might not be optimal. But this only affects the IMRPhenomXHM against the \bhpt case. 

\subsection{\label{sec:ff}Fitting Factors\protect}

\begin{figure*}[ht]
\begin{subfigure}{0.45\textwidth}
  \centering
  \includegraphics[width=\linewidth]{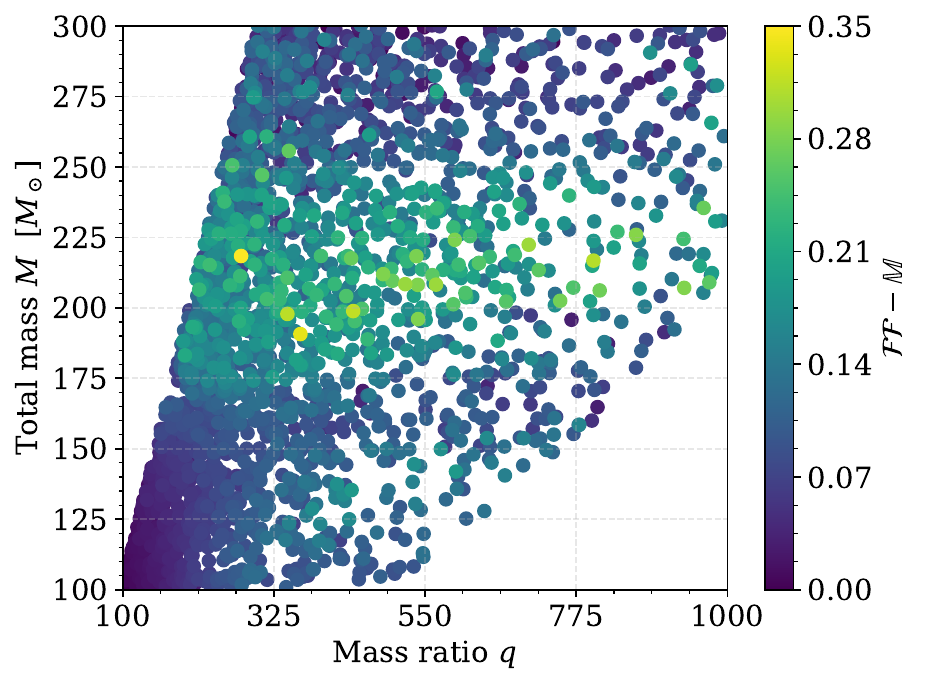}  
  \caption{$\mathcal{FF}-\mathbb{M}$ for $ \ell_\text{max}=5$}
  \label{fig:xas_all_ffm}
\end{subfigure}
\begin{subfigure}{0.45\textwidth}
  \centering
  \includegraphics[width=\linewidth]{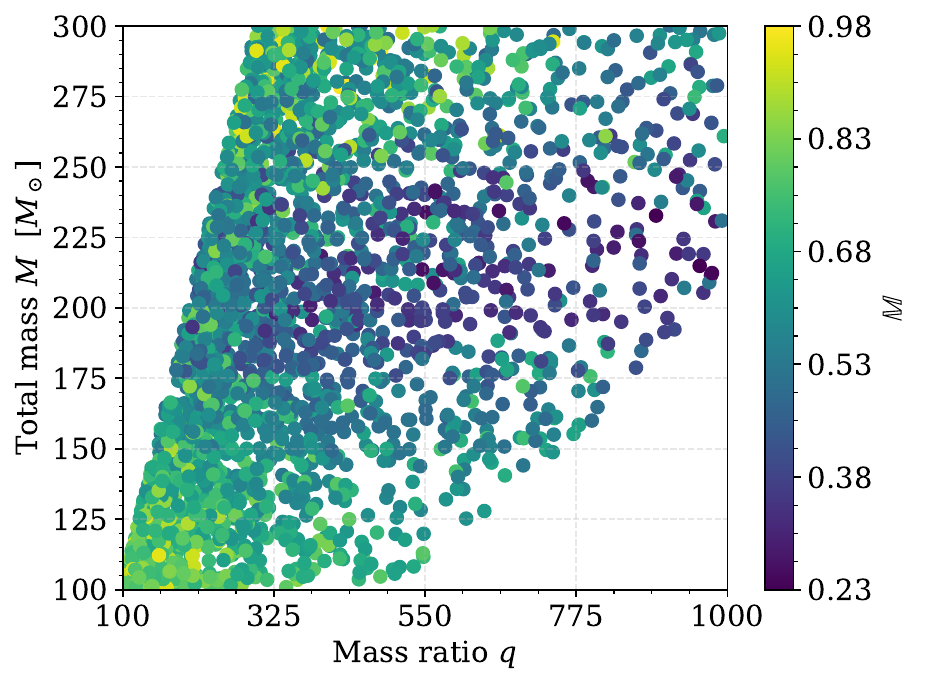}    
  \caption{$\mathbb{M}$ for $ \ell_\text{max}=5$}
  \label{fig:xas_all_m}
\end{subfigure}

\begin{subfigure}{0.45\textwidth}
  \centering
  \includegraphics[width=\linewidth]{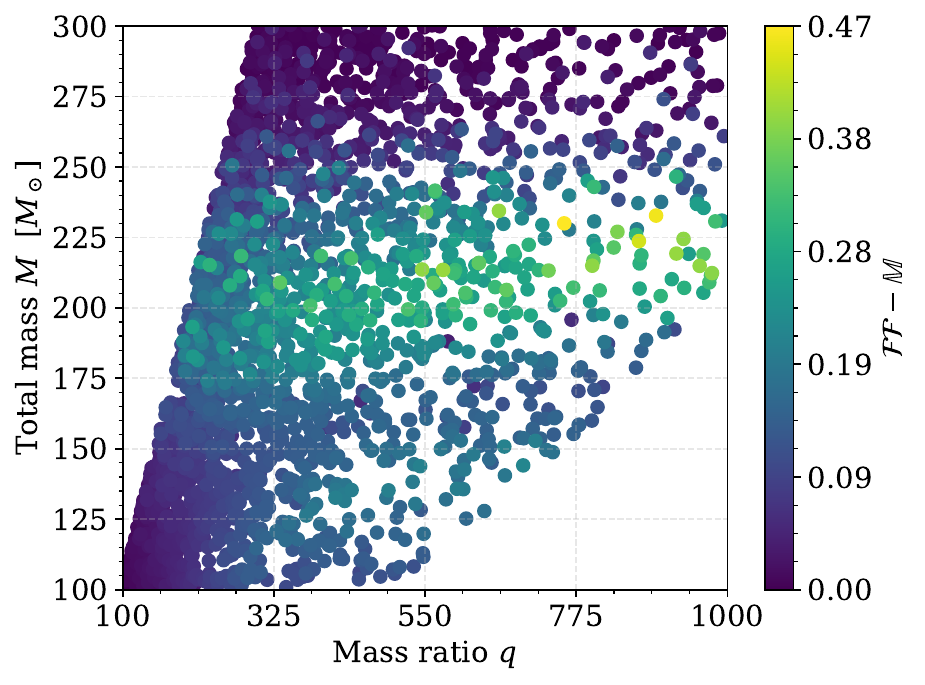}  
  \caption{$\mathcal{FF}-\mathbb{M}$ for $ \ell_\text{max}=2$}
  \label{fig:xas_22_ffm}
\end{subfigure}
\begin{subfigure}{0.45\textwidth}
  \centering
  \includegraphics[width=\linewidth]{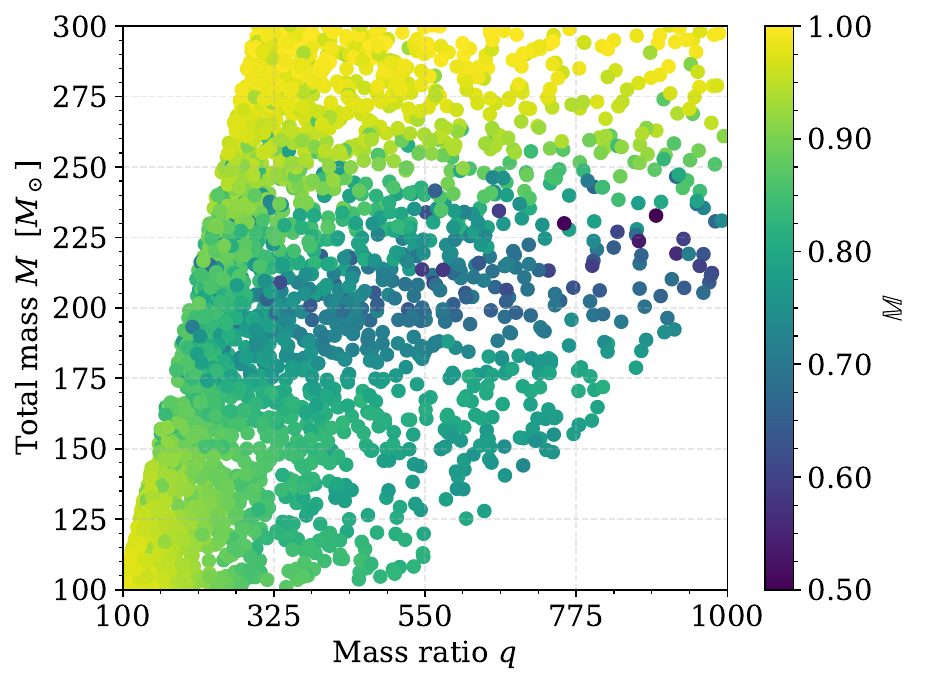}    
  \caption{$\mathbb{M}$ for $ \ell_\text{max}=2$}
  \label{fig:xas_22_m}
\end{subfigure}
\caption{\justifying{Fitting factor $\mathcal{FF}$ and match $\mathbb{M}$ in the $q-M$ plane for injections with \bhpt and recovery with \xas for aLIGO sensitivity and $f_\text{low} = 20$Hz. Fig. \ref{fig:xas_all_ffm} and \ref{fig:xas_all_m} show the difference $\mathcal{FF}-\mathbb{M}$ and match, respectively for injections using all the NR-calibrated modes up to $\ell_\text{max}=5$. Fig. \ref{fig:xas_22_ffm} and \ref{fig:xas_22_m} show the fitting factors and match, respectively, for injections using the NR-calibrated modes up to $\ell_\text{max}=2$.}}
\label{fig:ff_bhpt_xas}
\end{figure*}

\begin{figure*}[ht]
\begin{subfigure}{0.45\textwidth}
  \centering
  \includegraphics[width=\linewidth]{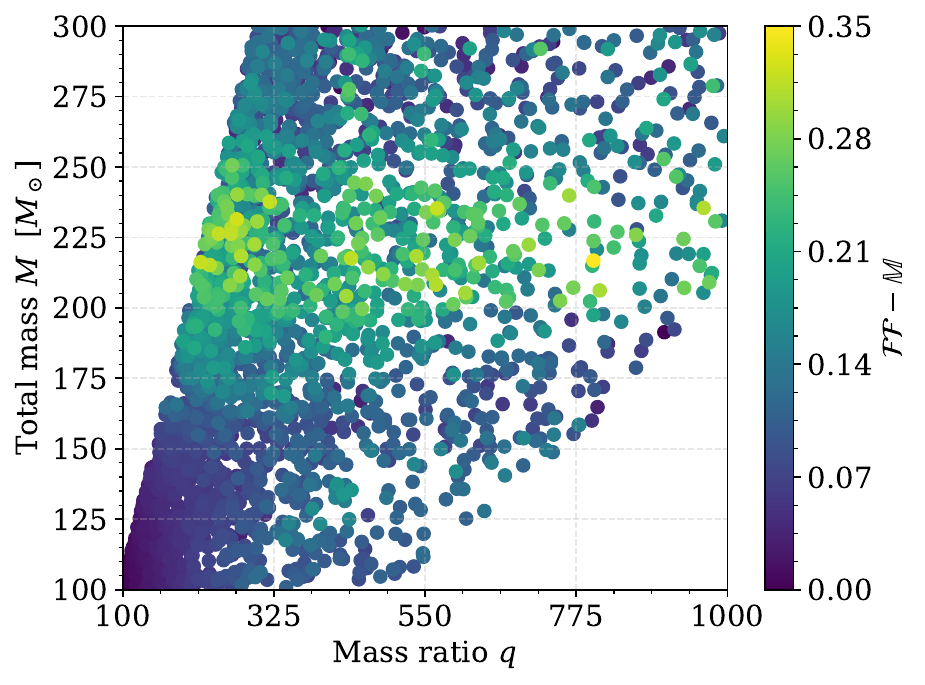}  
  \caption{$\mathcal{FF}-\mathbb{M}$ for $ \ell_\text{max}=5$}
  \label{fig:xhm_cal_ffm}
\end{subfigure}
\begin{subfigure}{0.45\textwidth}
  \centering
  \includegraphics[width=\linewidth]{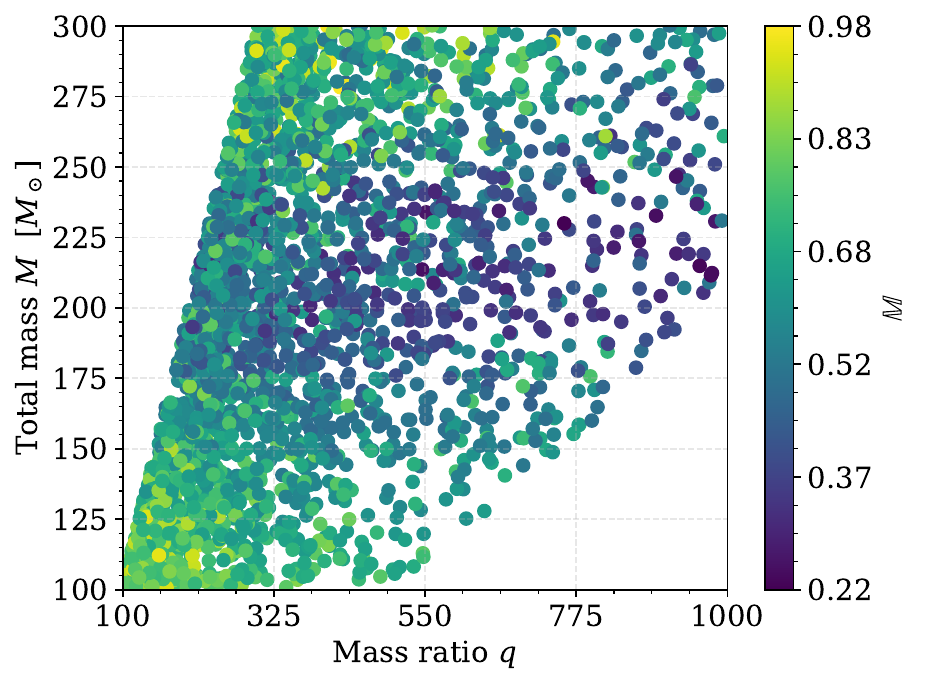}    
  \caption{$\mathbb{M}$ for $ \ell_\text{max}=5$}
  \label{fig:xhm_cal_m}
\end{subfigure}

\begin{subfigure}{0.45\textwidth}
  \centering
  \includegraphics[width=\linewidth]{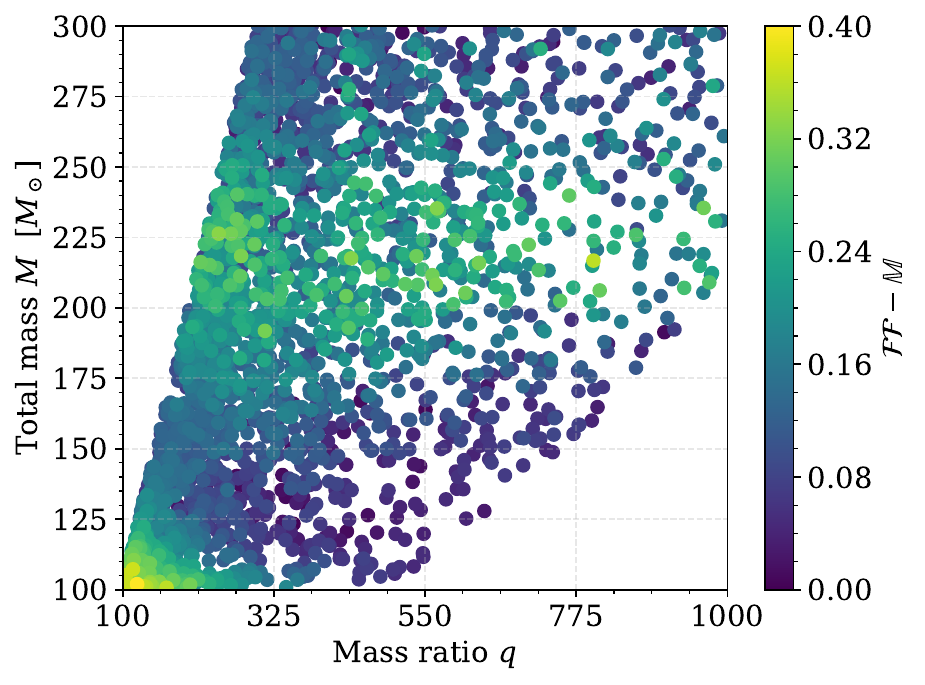}  
  \caption{$\mathcal{FF}-\mathbb{M}$ for $ \ell_\text{max}=10$}
  \label{fig:xhm_uncal_ffm}
\end{subfigure}
\begin{subfigure}{0.45\textwidth}
  \centering
  \includegraphics[width=\linewidth]{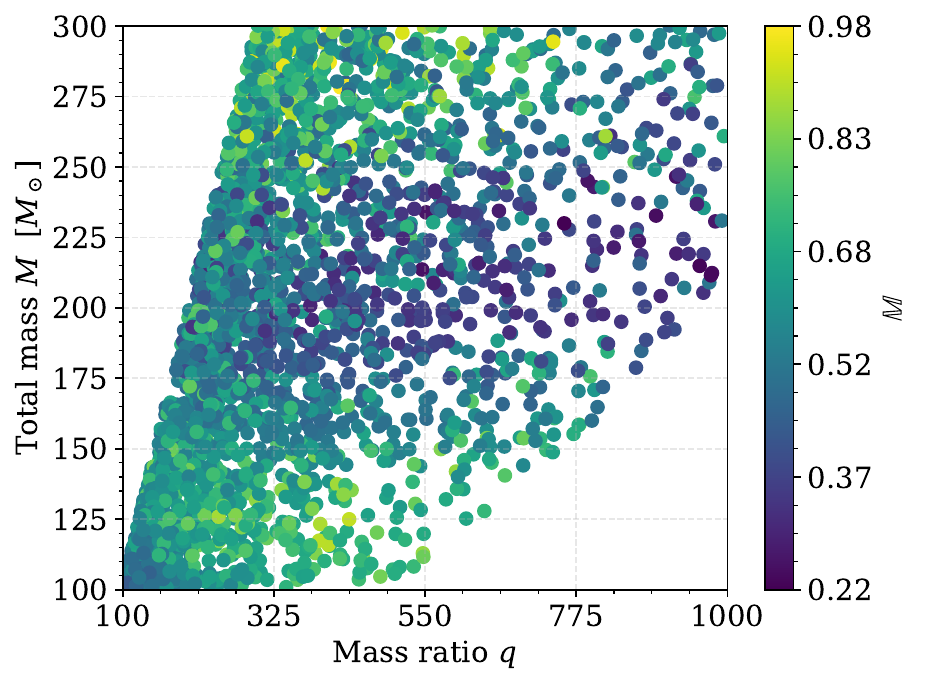}    
  \caption{$\mathbb{M}$ for $ \ell_\text{max}=10$}
  \label{fig:xhm_uncal_m}
\end{subfigure}
\caption{\justifying{Fitting factor $\mathcal{FF}$ and match $\mathbb{M}$ in the $q-M$ plane for injections with \bhpt and recovery with \xhm for aLIGO sensitivity and $f_\text{low} = 20$Hz. Fig. \ref{fig:xhm_cal_ffm} and \ref{fig:xhm_cal_m} show the difference $\mathcal{FF}-\mathbb{M}$ and match, respectively for injections using all the NR-calibrated modes up to $\ell_\text{max}=5$. Fig. \ref{fig:xhm_uncal_ffm} and \ref{fig:xhm_uncal_m} show the fitting factors and match, respectively, for injections using all the modes up to $\ell_\text{max}=10$ without NR-calibration.}}
\label{fig:ff_bhpt_xhm}
\end{figure*}

\begin{table*}[ht]
    \centering
    \caption{Summary of the fitting factor study. All the injections have been made assuming \bhpt as the true waveform. The first two columns give $\ell_\text{max}$ value and calibration settings for the injections. The third column mentions the waveform model that maximizes the match over masses and aligned spins. $\%_{\mathbb{M}\leq 0.9}$ shows the percentage of the total injected cases (i.e. $2580$) that have $\mathbb{M}\leq 0.9$. $\%_{\mathcal{FF}\leq 0.9}$ shows the percentage of total injected cases ($2580$) that have $\mathcal{FF}\leq 0.9$. $\%_{\mathcal{FF}\leq 0.9}^{\pi/4\leq \iota \leq 3\pi/4}$ shows the percentage of cases in the edge-on region $(\pi/4\leq \iota \leq 3\pi/4)$ that have $\mathcal{FF}\leq 0.9$. These results tabulated below are for aLIGO sensitivity and $f_\text{low}=20$Hz.} 
    \label{tab:ff_summary}
    \begin{tabular}{c|c|c|c|c|c}
        \hline
        NR-calibration & $\ell_\text{max}$ & Recovery & $\%_{\mathbb{M}\leq 0.9}$ & $\%_{\mathcal{FF}\leq 0.9}$ & $\%_{\mathcal{FF}\leq 0.9}^{\pi/4\leq \iota \leq 3\pi/4}$ \\
        \hline
        \hline
        \checkmark & $5$ & \xas & $97.64$ & $91.59$ & $100$ \\
        \hline
        \checkmark & $2$ & \xas & $59.26$ & $0.43$ & $0.54$ \\
        \hline
        \checkmark & $5$ & \xhm & $97.60$ & $90.97$ & $100$ \\
        \hline
        $\times$ & $10$ & \xhm & $98.60$ & $91.24$ & $100$ \\
        \hline
    \end{tabular}
\end{table*}

Fitting factor \cite{ff_intro} is a way to check the effectualness of GW waveform models. It quantifies how well the best template from a family of templates matches the actual waveform given a detector noise spectrum. The fitting factor, $\mathcal{FF}$, is defined as the best match (as defined in Eq. \ref{match_def}) between a normalized signal $s$ and normalized templates $h_i$ across the parameter space, i.e.,
\begin{align}
    \mathcal{FF} = \max_{m_1, m_2, s_{1z}, s_{2z}}{\mathbb{M}(s,h_i)} \label{ff_eq}
\end{align}

The fitting factors are a way to check the fraction of optimal SNR that can be obtained for a signal using a template signal, which may have systematic errors, making them suboptimal templates. Fitting factors are directly related to the horizon distance of the search, and the cube of fitting factors is related to the volume reach of the search \cite{Varma:2014jxa}.

\begin{figure*}[ht]
\begin{subfigure}{0.31\textwidth}
  \centering
  \includegraphics[width=\linewidth]{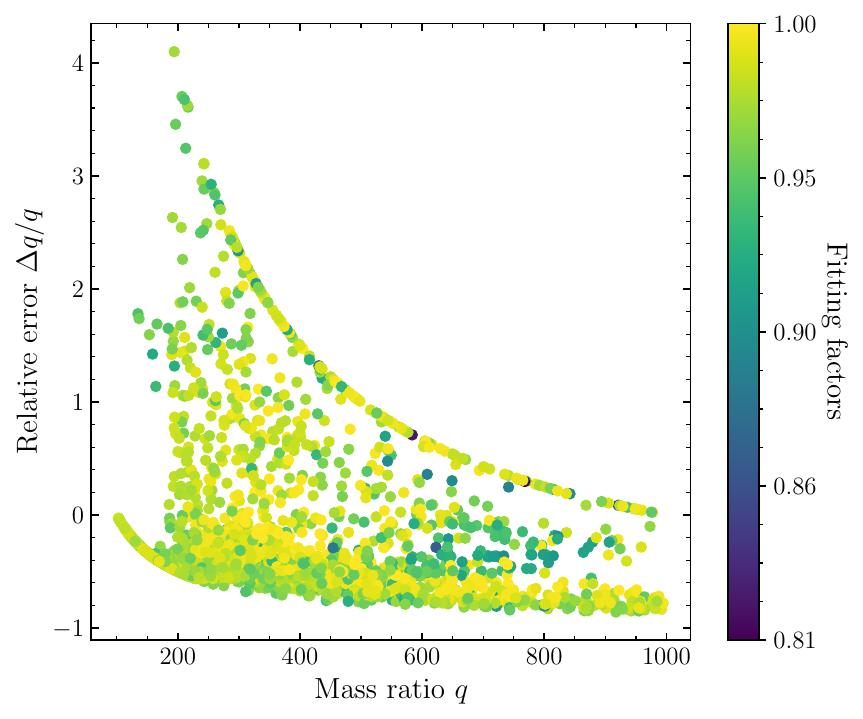}  
  \caption{\scriptsize{Injection: \textsc{\bhpt} $(2,\pm2)$ Recovery: \xas}}
  \label{fig:delta_q_by_q_xas_22}
\end{subfigure}
\begin{subfigure}{0.31\textwidth}
  \centering
  \includegraphics[width=\linewidth]{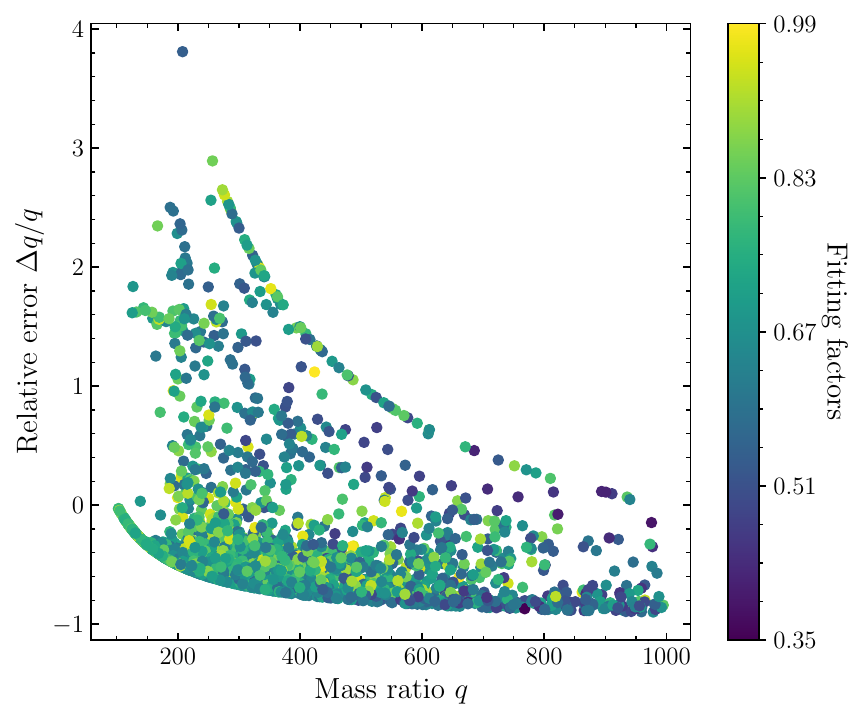}    
  \caption{\scriptsize{Injection: \textsc{\bhpt} (all) Recovery: \xas}}
  \label{fig:delta_q_by_q_xas_all}
\end{subfigure}
\begin{subfigure}{0.31\textwidth}
  \centering
  \includegraphics[width=\linewidth]{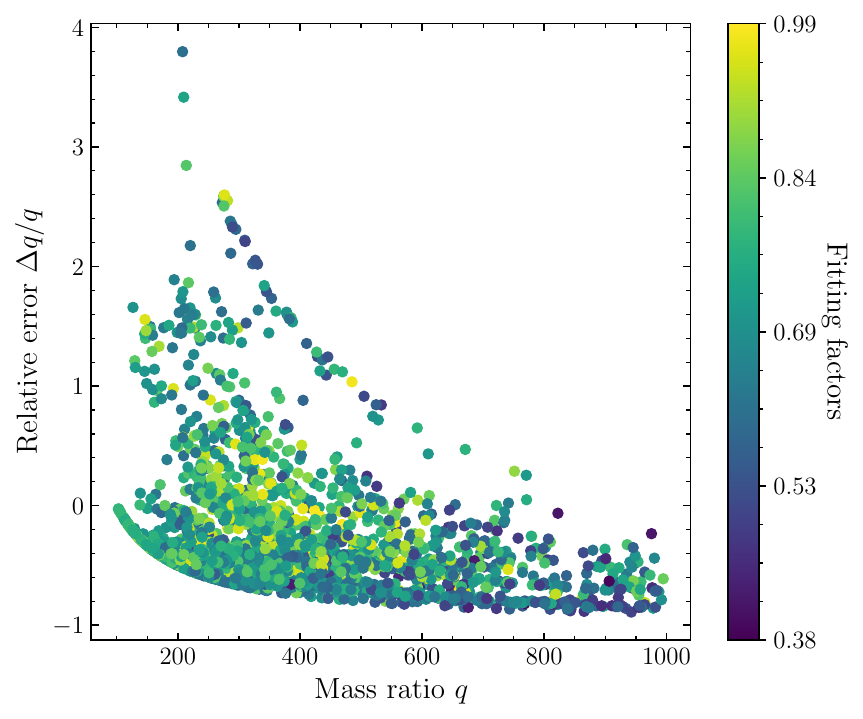}  
  \caption{\scriptsize{Injection: \textsc{\bhpt} (all) Recovery: \xhm}}
  \label{fig:delta_q_by_q_xhm_calibrated}
\end{subfigure}

\begin{subfigure}{0.31\textwidth}
  \centering
  \includegraphics[width=\linewidth]{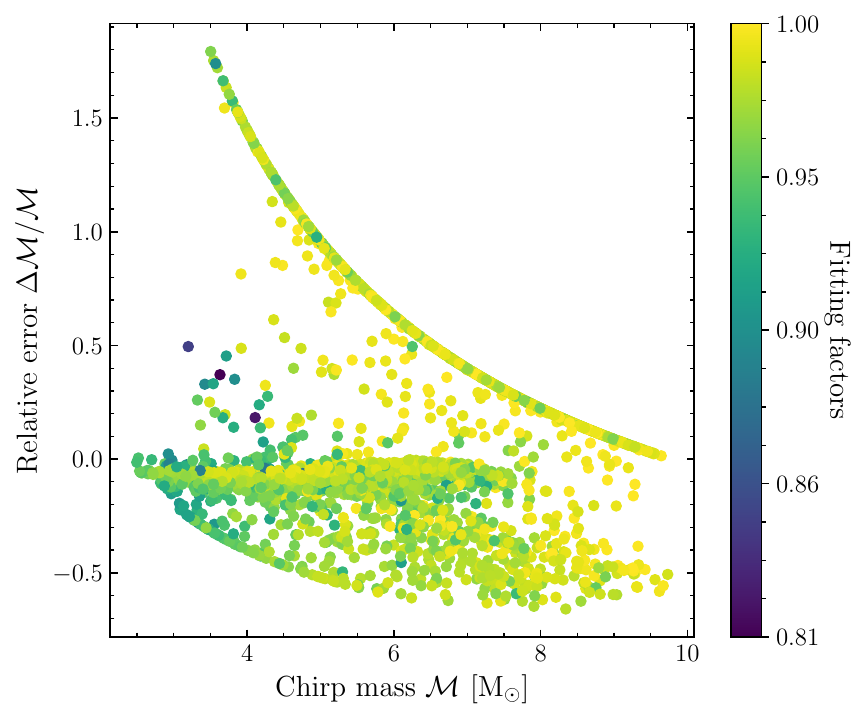}  
  \caption{\scriptsize{Injection: \textsc{\bhpt} $(2,\pm2)$ Recovery: \xas}}
  \label{fig:delta_mchirp_by_mchirp_xas_22}
\end{subfigure}
\begin{subfigure}{0.31\textwidth}
  \centering
  \includegraphics[width=\linewidth]{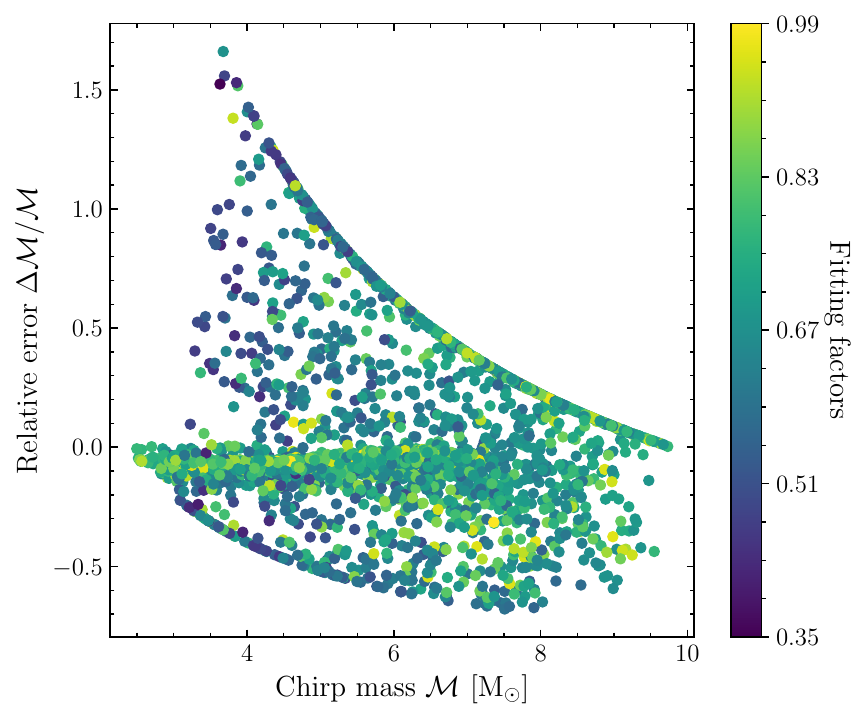}    
  \caption{\scriptsize{Injection: \textsc{\bhpt} (all) Recovery: \xas}}
  \label{fig:delta_mchirp_by_mchirp_xas_all}
\end{subfigure}
\begin{subfigure}{0.31\textwidth}
  \centering
  \includegraphics[width=\linewidth]{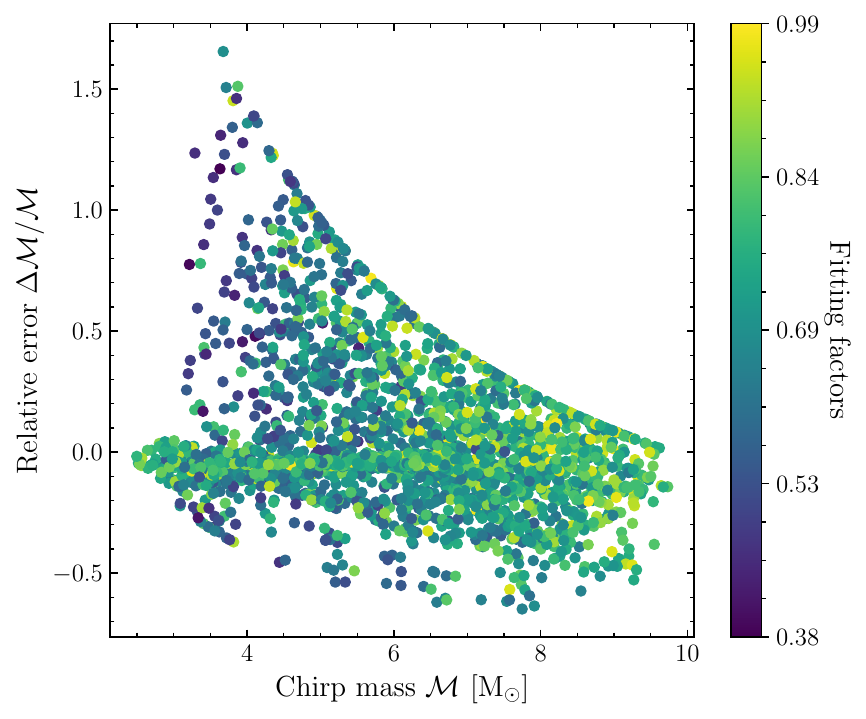}  
  \caption{\scriptsize{Injection: \textsc{\bhpt} (all) Recovery: \xhm}}
  \label{fig:delta_mchirp_by_mchirp_xhm_calibrated}
\end{subfigure}
\caption{\justifying{The above plots show the relative errors ${(k_\mathrm{rec}-k_\mathrm{inj})}/{k_\mathrm{inj}}$ in the recovery of the parameters through a fitting factor recovery scheme, where $k_\mathrm{inj}$ and $k_\mathrm{rec}$ are the injected and recovered values, respectively, of the parameter $k\in\{q,\mathcal{M}\}$. The left most column shows the results for the injections with only the NR-calibrated $(2,\pm2)$ modes of \bhpt. The next two columns show the results for the injections with all the NR-calibrated modes of \bhpt. The top panel shows the relative errors in mass ratio $q$, and the bottom panel shows the relative errors in chirp mass $\mathcal{M}$.}}
\label{fig:delta_k_by_k_ff_recovery}
\end{figure*}

\begin{figure*}[ht]
  \centering
  \includegraphics[width=0.99\textwidth]{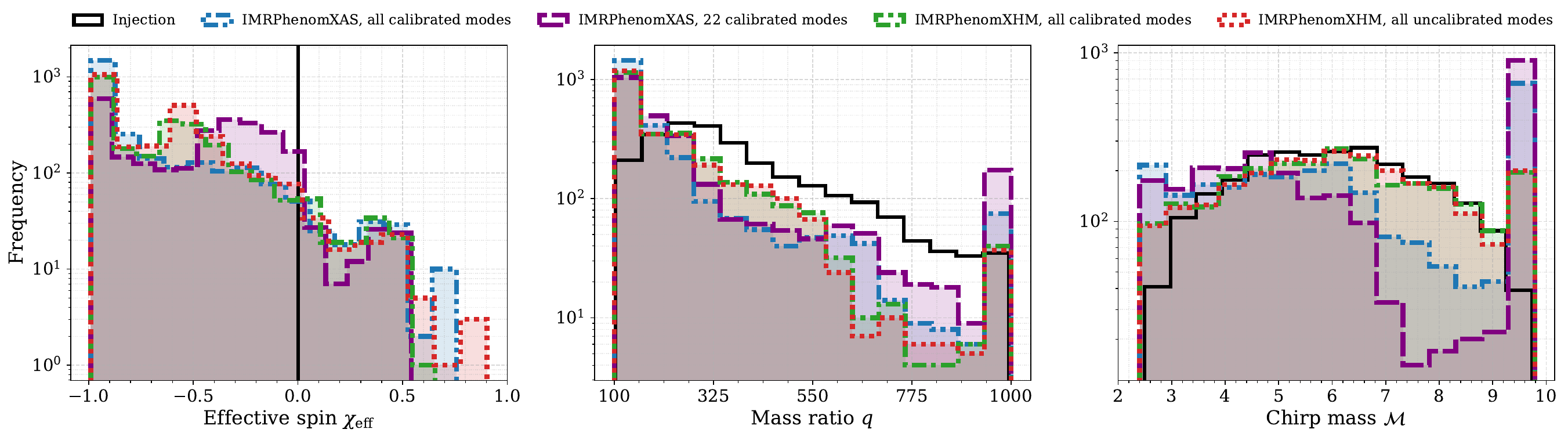}
  \caption{\justifying{Injected vs recovered population of SSM-IMRIs. The value of the parameters for which the fitting factor $\mathcal{FF}$ is obtained by differential evolution is quoted as the recovered value. The recovered population is shown for the parameters: effective spin $\chi_\text{eff}$ (left panel), mass ratio $q$ (center panel), and chirp mass $\mathcal{M}$ (right panel). The injected population is shown in gray. $\chi_\text{eff}=0$ for all injections. Different lines show population recovery with \xas or \xhm for injections made with the modes mentioned in the legend.}}
  \label{fig:pop_hist}
\end{figure*}

To find the best match template, we perform a continuous maximization in the mass-spin space using a global optimization technique, namely, differential evolution \cite{diff_evo_algo}. For this, we use \texttt{DiffEvol} implemented in the \texttt{SciPy} library \cite{scipy} with a population size of $20$ and the maximum allowed iterations as $500$. We make five separate runs with different seed values and pick the best of the matches as the fitting factor to avoid non-convergence problems in the differential evolution process. We perform a fitting factor study considering the \bhpt as the base model for injections and recovery using \xas and \xhm.
We inject $2580$ GW signals with: masses uniformly distributed in ranges $100M_\odot<m_1<300M_\odot$ and $0.2M_\odot<m_2<1M_\odot$ in the detector-frame with the constraint $100<m_1/m_2<1000$; inclination angle $\iota$ following a sine distribution 
in $[0,\pi]$; coalescence phase $\varphi_c$ and polarization $\psi$ distributed uniformly in $\left[0,2\pi\right)$; chirp distance 
distributed uniformly in $[4, 300]$ that gives a uniform volume prior; right ascension $\alpha$ and declination $\delta$ following a distribution uniform on the sky 
(i.e. uniform in solid angle) with $\alpha \in \left[0, 2\pi\right)$ and $\delta \in [-\pi/2, \pi/2]$. 
Of the total injections, $71.12\%$ cases lie in the edge-on region (i.e. $\pi/4 \leq \iota \leq 3\pi/4$).
We have already shown in Fig. \ref{fig:higher_mode_iq_plot} that the dependence of SNR content in higher modes for aLIGO depends significantly on inclination angle above $q=300$ which corresponds to $M\approx300\mathrm{M}_\odot$ given the constraint on $m_2$. Since we do not maximize over the inclination angles in the fitting factor recovery, we choose to consider the parameter space $M < 300\mathrm{M}_\odot$ to understand the systematic uncertainties in the parameter space of only the intrinsic parameters. The choice also stems from the CBC search perspective, which mostly uses the dominant mode only. Hence, the fitting factors also denote sensitivity lost due to the use of an inaccurate waveform approximant.
For recovery (i.e. for maximizing the match over masses and spins), we use chirp mass $\mathcal{M}$ and symmetric mass ratio $\eta$ as search parameters with $100\text{M}_\odot\leq m_1\leq 300\text{M}_\odot$, $0.2\text{M}_\odot\leq m_2\leq 1\text{M}_\odot$ and $100\leq q\leq 1000$. We also consider spins while recovering with $|s_{1z}|\leq 0.99$ and $|s_{2z}|\leq 0.99$. We have set the low-frequency cutoff as $20Hz$ and ensured a common overlap region in the frequency for all the waveforms. The maximum frequency for all the waveforms is set as $f_\text{max}=2048$Hz.

Fig. \ref{fig:ff_bhpt_xas} and \ref{fig:ff_bhpt_xhm} show the results of the fitting factor study. Fig. \ref{fig:xas_all_ffm} and \ref{fig:xas_22_ffm} show the difference between the fitting factors and the matches obtained for injections with \bhpt with NR-calibrated $\ell_\text{max}=5$ and $\ell_\text{max}=2$, respectively; and Fig. \ref{fig:xhm_cal_ffm} and \ref{fig:xhm_uncal_ffm} show the difference between the fitting factors and matches obtained for injections with \bhpt with NR-calibrated $\ell_\text{max}=5$ and NR-uncalibrated $\ell_\text{max}=10$. Fig. \ref{fig:xas_all_m}, \ref{fig:xas_22_m}, \ref{fig:xhm_cal_m} and \ref{fig:xhm_uncal_m} show the corresponding matches of the cases in the subplots on the left. $\mathcal{FF}-\mathbb{M}\geq 0$ since $\mathcal{FF}$ is obtained by maximizing $\mathbb{M}$ over masses and spins. 
In the region $175\text{M}_\odot\leq M \leq 250\text{M}_\odot$, we find that the bias is the highest for all the above-mentioned cases. The match $\mathbb{M}$ is also significantly poor in this region. 

For signals injected with \bhpt using $\ell_\text{max}=5$ (all NR-calibrated) and recovered with \xas, $97.64\%$ cases have $\mathbb{M}\leq 0.9$ while only $91.47\%$ cases have $\mathcal{FF}\leq 0.9$. All the signals in the edge-on region have $\mathcal{FF}\leq 0.9$, which makes $77.75\%$ of the total cases that have $\mathcal{FF}\leq 0.9$.

For signals injected with \bhpt using $\ell_\text{max}=2$ (all NR-calibrated) and recovered with \xas, $59.26\%$ cases have $\mathbb{M}\leq 0.9$ while only $0.43\%$ cases have $\mathcal{FF}\leq 0.9$. Only $0.54\%$ of the signals in the edge-on region have $\mathcal{FF}\leq 0.9$; however, this makes $90.91\%$ of the total cases that have $\mathcal{FF}\leq 0.9$.

For signals injected with \bhpt using $\ell_\text{max}=5$ (all NR-calibrated) and recovered with \xhm, $97.60\%$ cases have $\mathbb{M}\leq 0.9$ while only $90.97\%$ cases have $\mathcal{FF}\leq 0.9$. All signals in the edge-on region have $\mathcal{FF}\leq 0.9$, which makes $78.18\%$ of the total cases that have $\mathcal{FF}\leq 0.9$.

For signals injected with \bhpt using $\ell_\text{max}=10$ (all NR-uncalibrated) and recovered with \xhm, $98.60\%$ cases have $\mathbb{M}\leq 0.9$ while only $91.24\%$ cases have $\mathcal{FF}\leq 0.9$. All signals in the edge-on region have $\mathcal{FF}\leq 0.9$, which makes $77.95\%$ of the total cases that have $\mathcal{FF}\leq 0.9$. Table \ref{tab:ff_summary} gives a summary of the discussion above.

To understand the error in the recovery of the parameters across the parameter space through a fitting factor scheme, we show the relative error defined as $(k_\mathrm{inj}-k_\mathrm{rec})/k_\mathrm{inj}$ where $k_\mathrm{inj}$ and $k_\mathrm{rec}$ denote the injected and recovered parameter values for the parameter $k$. The results for the relative errors in mass ratio and chirp mass are shown in Fig. \ref{fig:delta_k_by_k_ff_recovery}. We find that the mass ratio and chirp mass recovery estimates can be biased to as high as $\sim5$ and $\sim3$ times, respectively.

Fig. \ref{fig:pop_hist} shows the population recovery for the injected SSM-IMRI signals used for the fitting factor study. The parameters for which the $\mathcal{FF}$ values are obtained using differential evolution are the recovered values for each signal. The solid blue lines show the recovery using \xas for injections with all the NR-calibrated modes of \bhpt. The orange dashed lines show the recovery using \xas for injections with only the $(2,\pm 2)$ NR-calibrated modes of \bhpt. The green dash-dot lines show the recovery using \xhm for injections with all the NR-calibrated modes of \bhpt. The red dotted lines show the recovery using \xhm for injections with all NR-uncalibrated modes of \bhpt. The injected population is shown in grey (solid patch). In Fig. \ref{fig:pop_hist}: the left panel shows the population of the recovered effective spin $\chi_\text{eff}$; the centre panel shows the population of the recovered mass ratio $q$; the right panel shows the population of the recovered chirp mass $\mathcal{M}$.
These plots roughly correspond to the expected bias we may get in the population analysis due to systematic errors in the waveforms used in the analysis. 

\subsection{\label{sec:pe}Parameter Estimation\protect}

Given the GW strain data $d(t)$ and a theoretical model (or hypothesis for the data, here, general relativity) $\mathcal{H}$, we can infer the source parameters $\boldsymbol{\theta}$ of gravitational waves from CBCs using Bayesian parameter estimation (PE) based on Bayes' theorem which states
\begin{align}
    p(\boldsymbol{\theta}|d,\mathcal{H}) = \frac{\mathcal{L}(d|\boldsymbol{\theta},\mathcal{H})\pi(\boldsymbol{\theta}|\mathcal{H})}{\mathcal{Z}(d|\mathcal{H})}.
\end{align}
The quantity $p(\boldsymbol{\theta}|d,\mathcal{H})$ is called the \textit{posterior probability distribution} which is the probability density of $\boldsymbol{\theta}$ given $d$ and $\mathcal{H}$; $\mathcal{L}(d|\boldsymbol{\theta},\mathcal{H})$ is called the \textit{likelihood} which is the probability density of $d$ given $\boldsymbol{\theta}$ and $\mathcal{H}$; $\pi(\boldsymbol{\theta}|\mathcal{H})$ is called the \textit{prior} which is the probability density of the parameters $\boldsymbol{\theta}$ given the model $\mathcal{H}$; and $\mathcal{Z}(d|\mathcal{H})$ is called the \textit{evidence} which is the marginalized likelihood of $d$ given a model $\mathcal{H}$. Generally, $\boldsymbol{\theta}$ represents a $15$-dimensional parameter space of circular binary black hole mergers. 

We describe the Bayesian PE for our study of non-spinning SSM-IMRI systems. We consider two injections with primary mass to $100 \textrm{M}_\odot$ and mass ratios of 250 and 500. This fixes secondary masses to be $0.4\textrm{M}_\odot$ and $0.2\textrm{M}_\odot$ respectively, which are within the search limits of current SSM searches~\cite{ssm1, ssm3}. We use NR-calibrated \bhpt as injections and recover using \xas and \xhm. We fix the inclination angle at $\iota = \pi/3$ and the minimum frequency at $20$ Hz for our analyses, considering a three-detector network comprising the LIGO Hanford, LIGO Livingston, and Virgo detectors at their design sensitivities. The luminosity distance is fixed by choosing the injected SNR to be $\approx25$. For $q=250$ and $500$ cases, the luminosity distances are 64.18 Mpc and 39.77 Mpc. We use uniform priors on mass ratio and chirp mass with constraints on the primary and secondary mass such that $50\mathrm{M}_\odot\leq m_1\leq 150\mathrm{M}_\odot$ and $0.2\mathrm{M}_\odot\leq m_2\leq 1\mathrm{M}_\odot$. The prior for luminosity distance is set to be uniform in commoving volume and source frame time, and the prior for inclination angle is set to be uniform in the sine of inclination angle.

\begin{table}[ht]
    \centering
    \centering{\textit{Mass ratio: 250}}
    \\
    \vspace{2pt}
    \begin{tabular}{|c|c|c|c|c|}
         \hline
         \textbf{Recovery model} & $\mathbf{\Delta M/\boldsymbol{\sigma_M}}$ & $\mathbf{\Delta q/\boldsymbol{\sigma_q}}$ & $\mathbf{\Delta d_\text{L}/\boldsymbol{\sigma_{d_L}}}$ & $\mathbf{\Delta \boldsymbol{\iota}/\boldsymbol{\sigma_\iota}}$ \\
         \hline
         \xas & $-12.32$ & $-13.19$ & $1.70$ & $-2.00$ \\
         \xhm & $11.90$ & $10.08$ & $2.61$ & $-0.70$ \\
         \hline
    \end{tabular}
    \\
    \vspace{5pt}
    \centering{\textit{Mass ratio: 500}}
    \\
    \vspace{2pt}
    \begin{tabular}{|c|c|c|c|c|}
         \hline
         \textbf{Recovery model} & $\mathbf{\Delta M/\boldsymbol{\sigma_M}}$ & $\mathbf{\Delta q/\boldsymbol{\sigma_q}}$ & $\mathbf{\Delta d_\text{L}/\boldsymbol{\sigma_{d_L}}}$ & $\mathbf{\Delta \boldsymbol{\iota}/\boldsymbol{\sigma_\iota}}$ \\
         \hline
         \xas & $-12.16$ & $-8.31$ & $1.82$ & $-1.43$ \\
         \xhm & $-5.34$ & $-4.65$ & $6.04$ & $-0.79$ \\
         \hline
    \end{tabular}
    \caption{\justifying{The above table shows the relative errors in the recovery of the parameters for injections with \bhpt considering $\ell_\text{max}=5$. The top and bottom panels show the results for the mass ratio cases of 250 and 500, respectively.}}
    \label{tab:pe_table}
\end{table}

\begin{figure*}[htp!]
\begin{subfigure}{\textwidth}
  \centering
  \includegraphics[width=0.6\textwidth]{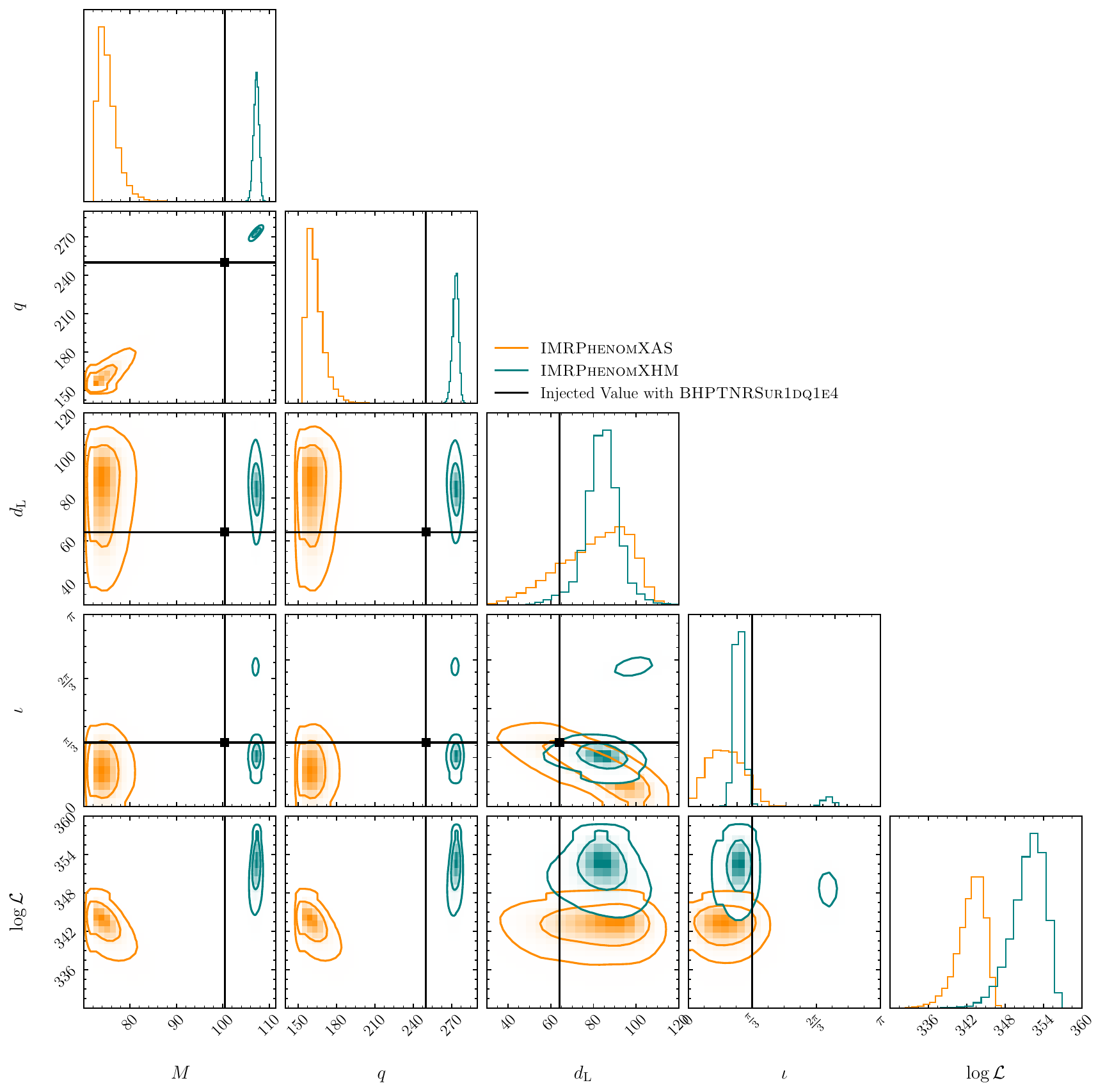}
  \caption{Mass ratio: 250}
  \label{fig:pe250}
\end{subfigure}
\begin{subfigure}{\textwidth}
  \centering
  \includegraphics[width=0.6\textwidth]{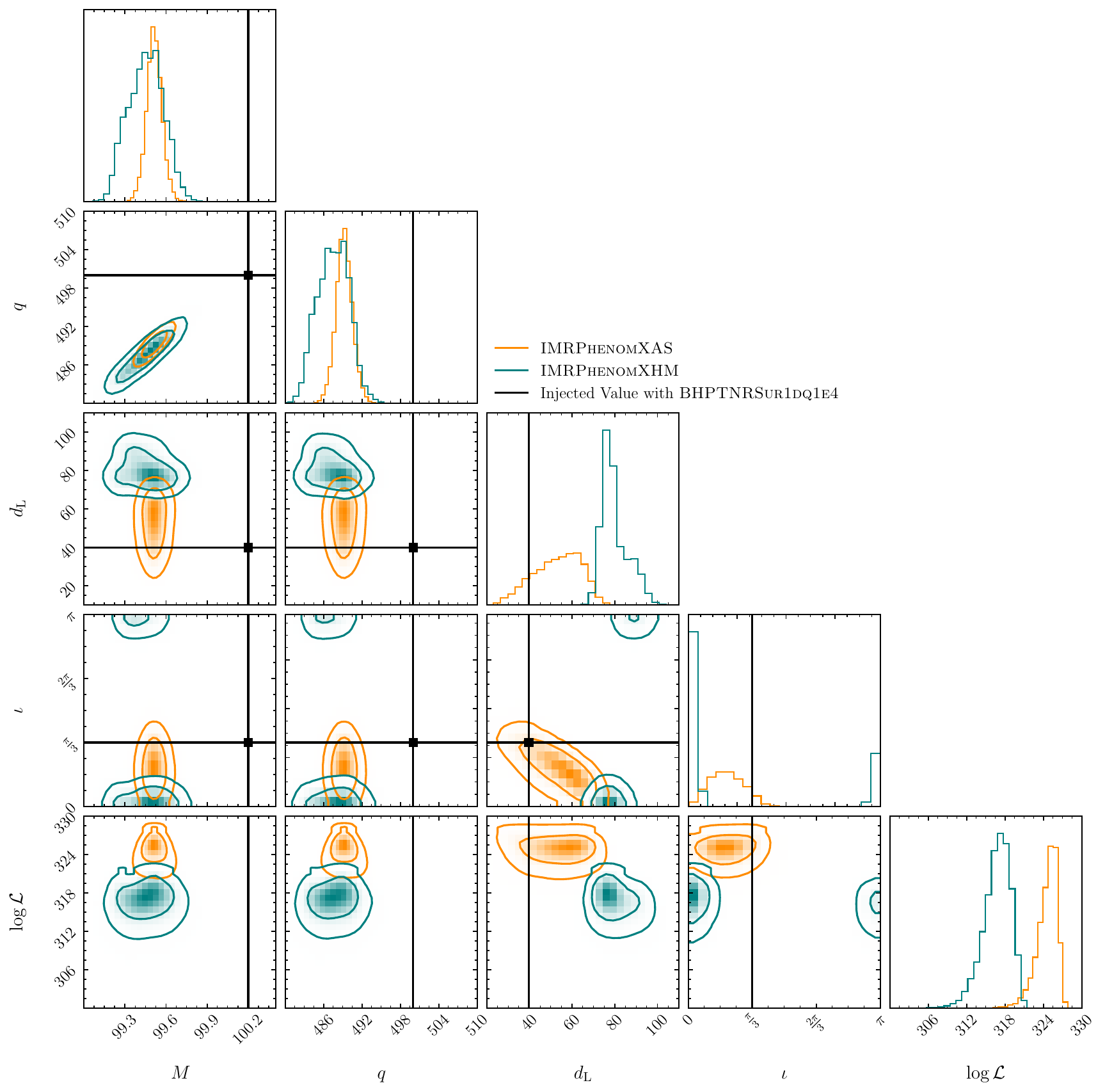}
  \caption{Mass ratio: 500}
  \label{fig:pe500}
\end{subfigure}
\caption{\justifying{Posterior distributions for total mass $M$, mass ratio $q$, luminosity distance $d_\mathrm{L}$, and the angle of inclination $\iota$. The injected values are represented by black solid lines. The contours show the $95\%$ and $68\%$ credible regions in the 2D corner plots. Contours of teal and orange colors denote recovery using \xhm and \xas waveform models, respectively.}}
\label{fig:pe}
\end{figure*}

\begin{figure}[ht]
    \centering
    \includegraphics[width=\linewidth]{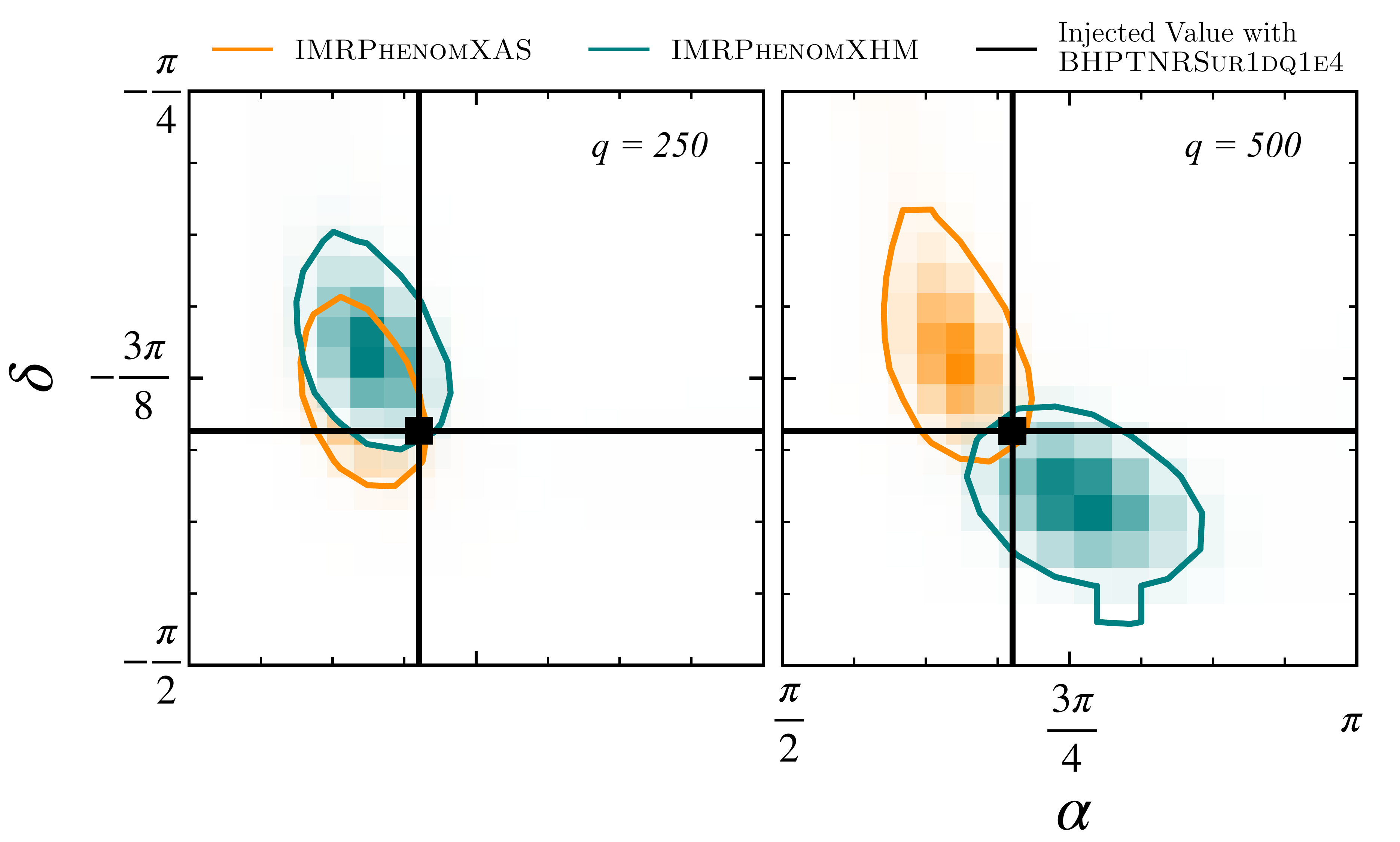}
    \caption{\justifying{The above plot shows a comparison of the posterior distributions for right ascension $(\alpha)$ and declination $(\delta)$ for both the mass ratio cases of $q=250$ (left panel) and $q=500$ (right panel).}}
    \label{fig:ra_dec_combined}
\end{figure}

The posteriors for total mass $M$, mass ratio $q$, luminosity distance $d_\mathrm{L}$, and the angle of inclination $\iota$ along with the likelihood values are shown in Figure \ref{fig:pe}. In general, posteriors completely miss the injected total mass and mass ratio for both cases. For the q=250 case, we can see an overall improvement in parameter recovery when using \xhm over \xas due to the availability of higher harmonic content. Interestingly, the \xas posterior prefers a 24\% lower total mass compared to \xhm, which converges on a 10\% larger total mass compared to the injected value. The deviation of the quadrupolar mode and differences in higher mode content of \textsc{IMRPhenomX} models compared to the \bhpt waveform increase with the increasing mass ratio, as evident from the fitting factor study. However, we observe consistent total mass recovery between the two \textsc{IMRPhenoX} models for the q=500 case, which is 10\% lower than the injected value. The same is true for the mass ratio recovery as well. We do not have a clear explanation for this recovery; however, the lower likelihood values associated with posteriors for the q=500 case, compared to the q=250 case, make it evident that waveform models diverge systematically with mass ratio, as observed before.  We quantify the relative errors on the recovery of the parameters by $(k_\mathrm{MAP}-k_\mathrm{inj})/\sigma_k$ where $k_\mathrm{inj}$ denotes the injected value, $k_\mathrm{MAP}$ denotes the \textit{maximum a posteriori} estimate, and $\sigma_k$ denotes the standard deviation of the 1D marginalized posterior for the parameter $k$. The relative errors on the parameter estimation results when recovered using \xas and \xhm for injections with \bhpt (all NR-calibrated modes) are summarized in Table \ref{tab:pe_table}.

We also compare the sky-localization of SSM-IMRI systems for both the mass ratio cases and the relative error when recovering using the \xas and \xhm waveforms. We found that for $q=250$ case, the relative errors in right ascension-declination $(\alpha,\delta)$ are $(-1.58,\,0.20)$ and $(-0.57,\,0.87)$ for recovery with \xas and \xhm, respectively. For $q=500$ case, the corresponding relative errors in $(\alpha,\delta)$ are $(-1.41,\,0.61)$ and $(0.23,\,-1.31)$ for \xas and \xhm, respectively. But in general, sky localization is not largely affected by the presence of waveform systematics.

\section{\label{sec:conclusions}Conclusion\protect}

We have studied the detectability and waveform systematics of sub-solar mass intermediate mass-ratio inspirals (SSM-IMRIs) by considering the state-of-the-art \bhpt as our reference (true) waveform model and comparing it against the \textsc{IMRPhenomX} family of phenomenological models which includes \xas and \xhm waveforms. The extreme asymmetry in the masses of these systems $(q \sim 10^2-10^3)$ places these systems beyond the traditional calibration range of most inspiral-merger-ringdown (IMR) models, making waveform model selection a decisive factor in both: gravitational wave search effectualness and parameter estimation. Our analyses show that although \textsc{IMRPhenomX} family waveforms reproduce the gross morphology of the dominant $(2,\pm2)$ quadrupole mode of GWs, they fail to capture the amplitude and phase evolution of the higher harmonics that contribute significantly at high values of mass ratios and edge-on inclination angles. While \bhpt is expected to represent the correct dynamics of SSM-IMRIs, the waveform generation is extremely slow and, combined with limited waveform duration, makes \bhpt unattractive for template-bank-based searches.  On the other hand, \textsc{IMRPhenomX} waveform models are fast but fail to reproduce the accuracy of \bhpt. The resulting systematic discrepancies directly impact detection statistics and the accuracy of source parameter inference.

Using matched-filter calculations considering the design sensitivities of the current and future generations of GW detectors, we find that SSM-IMRIs are, in principle, observable by Advanced LIGO out to $\sim575$ MPc $(z\sim 0.12)$ and by the Einstein Telescope out to $\sim10.5$ Gpc $(z\sim1.4)$ when higher modes in GWs are included. Inclusion of higher harmonics can enhance the SNR by factors of $\sim 3-5$ relative to the quadrupole-only templates, thereby increasing the detection volume by more than an order of magnitude. These effects become increasingly significant for edge-on orientations of binaries. These results demonstrate that higher modes of GWs are essential for any realistic GW search strategy targeting IMRIs in general, and SSM-IMRIs in particular.

A detailed comparison between \bhpt and \textsc{IMRPhenomX} family reveals substantial waveform disagreement that is increasing with the mass asymmetry. The matches between the waveforms can degrade to less than 40\% for edge-on sources. All edge-on configurations yield fitting factors below 0.9, implying that the detection and unbiased characterization of such sources is challenging. Parameter estimation runs confirm that these modeling discrepancies propagate into strong systematic biases exceeding statistical uncertainties by multiple standard deviations in source parameters, and also produce unreliable inference on the source geometry and possibly source localisation in the sky. All these results strongly suggest that current phenomenological models, which are fast to generate, can not be extrapolated to large mass ratios for robust detection and astrophysical inference of IMRIs.

Realizing the full potential of SSM-IMRIs requires accurate and faster waveform models that combine the accuracy of black-hole perturbation theory with the computational efficiency of phenomenological model frameworks, incorporating calibrated higher-modes and spin-coupling effects across $q>100$. From the lens of gravitational wave data analysis, search pipelines must be adapted to utilize such models, either through multi-mode template banks or hierarchical follow-ups using mode-by-mode template waveforms to recover these systems efficiently. With these developments, SSM-IMRIs could become precision probes of the low-mass black hole population and early-universe compact-object formation, along with intermediate-mass black holes, both of which remain enigmatic to date. Until then, interpretations of such systems based on the current waveform scenario should be regarded as limited systematically. \\

\section{\label{sec:acknow}Acknowledgments\protect}

DG greatly acknowledges discussions with and suggestions from Sarah Caudill, and Stefano Schmidt. The authors are grateful for the computational resources provided by Nikhef, NWO. The authors acknowledge the use of the following software packages in the work: \textsc{PyCBC} ~\cite{PyCBC}, \textsc{SciPy} ~\cite{scipy}, \textsc{NumPy} ~\cite{numpy}, \textsc{Bilby} ~\cite{bilby_paper}, and \textsc{Matplotlib} ~\cite{matplotlib}.

\bibliography{apssamp}

@article{Hawking_collapse,
    author = "Hawking, Stephen",
    title = "{Gravitationally collapsed objects of very low mass}",
    doi = "10.1093/mnras/152.1.75",
    journal = "Mon. Not. Roy. Astron. Soc.",
    volume = "152",
    pages = "75",
    year = "1971"
}

@article{Timmes:1995kp,
    author = "Timmes, F. X. and Woosley, S. E. and Weaver, Thomas A.",
    title = "{The Neutron star and black hole initial mass function}",
    eprint = "astro-ph/9510136",
    archivePrefix = "arXiv",
    doi = "10.1086/176778",
    journal = "Astrophys. J.",
    volume = "457",
    pages = "834",
    year = "1996"
}

@article{Suwa:2018uni,
    author = "Suwa, Yudai and Yoshida, Takashi and Shibata, Masaru and Umeda, Hideyuki and Takahashi, Koh",
    title = "{On the minimum mass of neutron stars}",
    eprint = "1808.02328",
    archivePrefix = "arXiv",
    primaryClass = "astro-ph.HE",
    reportNumber = "YITP-18-98",
    doi = "10.1093/mnras/sty2460",
    journal = "Mon. Not. Roy. Astron. Soc.",
    volume = "481",
    number = "3",
    pages = "3305--3312",
    year = "2018"
}

@article{Carr:2020gox,
    author = "Carr, Bernard and Kohri, Kazunori and Sendouda, Yuuiti and Yokoyama, Jun'ichi",
    title = "{Constraints on primordial black holes}",
    eprint = "2002.12778",
    archivePrefix = "arXiv",
    primaryClass = "astro-ph.CO",
    reportNumber = "RESCEU-03/20; KEK-Cosmo-249; KEK-TH-2199; IPMU20-0024",
    doi = "10.1088/1361-6633/ac1e31",
    journal = "Rept. Prog. Phys.",
    volume = "84",
    number = "11",
    pages = "116902",
    year = "2021"
}

@article{Sasaki:2018dmp,
    author = "Sasaki, Misao and Suyama, Teruaki and Tanaka, Takahiro and Yokoyama, Shuichiro",
    title = "{Primordial black holes\textemdash{}perspectives in gravitational wave astronomy}",
    eprint = "1801.05235",
    archivePrefix = "arXiv",
    primaryClass = "astro-ph.CO",
    doi = "10.1088/1361-6382/aaa7b4",
    journal = "Class. Quant. Grav.",
    volume = "35",
    number = "6",
    pages = "063001",
    year = "2018"
}

@article{Green:2020jor,
    author = "Green, Anne M. and Kavanagh, Bradley J.",
    title = "{Primordial Black Holes as a dark matter candidate}",
    eprint = "2007.10722",
    archivePrefix = "arXiv",
    primaryClass = "astro-ph.CO",
    doi = "10.1088/1361-6471/abc534",
    journal = "J. Phys. G",
    volume = "48",
    number = "4",
    pages = "043001",
    year = "2021"
}

@article{Shandera:2018xkn,
    author = "Shandera, Sarah and Jeong, Donghui and Gebhardt, Henry S. Grasshorn",
    title = "{Gravitational Waves from Binary Mergers of Subsolar Mass Dark Black Holes}",
    eprint = "1802.08206",
    archivePrefix = "arXiv",
    primaryClass = "astro-ph.CO",
    reportNumber = "IGC-18-2-1",
    doi = "10.1103/PhysRevLett.120.241102",
    journal = "Phys. Rev. Lett.",
    volume = "120",
    number = "24",
    pages = "241102",
    year = "2018"
}

@article{Kouvaris:2018wnh,
    author = "Kouvaris, Chris and Tinyakov, Peter and Tytgat, Michel H. G.",
    title = "{NonPrimordial Solar Mass Black Holes}",
    eprint = "1804.06740",
    archivePrefix = "arXiv",
    primaryClass = "astro-ph.HE",
    reportNumber = "ULB-TH-18-05, CP3-ORIGINS-2018-014, ULB-TH/18-05, CP3-Origins-2018-014 DNRF90",
    doi = "10.1103/PhysRevLett.121.221102",
    journal = "Phys. Rev. Lett.",
    volume = "121",
    number = "22",
    pages = "221102",
    year = "2018"
}

@article{Clesse:2017bsw,
    author = "Clesse, Sebastien and Garc\'\i{}a-Bellido, Juan",
    title = "{Seven Hints for Primordial Black Hole Dark Matter}",
    eprint = "1711.10458",
    archivePrefix = "arXiv",
    primaryClass = "astro-ph.CO",
    reportNumber = "IFT-UAM-CSIC-17-108, CERN-TH-2017-239",
    doi = "10.1016/j.dark.2018.08.004",
    journal = "Phys. Dark Univ.",
    volume = "22",
    pages = "137--146",
    year = "2018"
}

@article{LIGO,
    author = "Aasi, J. and others",
    collaboration = "LIGO Scientific",
    title = "{Advanced LIGO}",
    eprint = "1411.4547",
    archivePrefix = "arXiv",
    primaryClass = "gr-qc",
    doi = "10.1088/0264-9381/32/7/074001",
    journal = "Class. Quant. Grav.",
    volume = "32",
    pages = "074001",
    year = "2015"
}

@article{VIRGO,
    author = "Acernese, F. and others",
    collaboration = "VIRGO",
    title = "{Advanced Virgo: a second-generation interferometric gravitational wave detector}",
    eprint = "1408.3978",
    archivePrefix = "arXiv",
    primaryClass = "gr-qc",
    doi = "10.1088/0264-9381/32/2/024001",
    journal = "Class. Quant. Grav.",
    volume = "32",
    number = "2",
    pages = "024001",
    year = "2015"
}

@article{KAGRA:2020tym,
    author = "Akutsu, T. and others",
    collaboration = "KAGRA",
    title = "{Overview of KAGRA: Detector design and construction history}",
    eprint = "2005.05574",
    archivePrefix = "arXiv",
    primaryClass = "physics.ins-det",
    doi = "10.1093/ptep/ptaa125",
    journal = "PTEP",
    volume = "2021",
    number = "5",
    pages = "05A101",
    year = "2021"
}

@article{Blanchet:2013haa,
    author = "Blanchet, Luc",
    title = "{Post-Newtonian Theory for Gravitational Waves}",
    eprint = "1310.1528",
    archivePrefix = "arXiv",
    primaryClass = "gr-qc",
    doi = "10.12942/lrr-2014-2",
    journal = "Living Rev. Rel.",
    volume = "17",
    pages = "2",
    year = "2014"
}

@article{Finn:1992xs,
    author = "Finn, Lee Samuel and Chernoff, David F.",
    title = "{Observing binary inspiral in gravitational radiation: One interferometer}",
    eprint = "gr-qc/9301003",
    archivePrefix = "arXiv",
    reportNumber = "PRINT-93-0138 (NORTHWESTERN)",
    doi = "10.1103/PhysRevD.47.2198",
    journal = "Phys. Rev. D",
    volume = "47",
    pages = "2198--2219",
    year = "1993"
}

@phdthesis{Li:2013lza,
    author = "Li, Tjonnie Guang Feng",
    title = "{Extracting Physics from Gravitational Waves: Testing the Strong-field Dynamics of General Relativity and Inferring the Large-scale Structure of the Universe}",
    school = "Vrije U., Amsterdam, Vrije U., Amsterdam",
    year = "2013",
    url = "{https://link.springer.com/book/10.1007/978-3-319-19273-4}"
}

@article{IMRPhenomXAS,
    author = "Pratten, Geraint and Husa, Sascha and Garcia-Quiros, Cecilio and Colleoni, Marta and Ramos-Buades, Antoni and Estelles, Hector and Jaume, Rafel",
    title = "{Setting the cornerstone for a family of models for gravitational waves from compact binaries: The dominant harmonic for nonprecessing quasicircular black holes}",
    eprint = "2001.11412",
    archivePrefix = "arXiv",
    primaryClass = "gr-qc",
    reportNumber = "LIGO-P2000018",
    doi = "10.1103/PhysRevD.102.064001",
    journal = "Phys. Rev. D",
    volume = "102",
    number = "6",
    pages = "064001",
    year = "2020"
}

@article{IMRPhenomXHM,
    author = "Garc\'\i{}a-Quir\'os, Cecilio and Colleoni, Marta and Husa, Sascha and Estell\'es, H\'ector and Pratten, Geraint and Ramos-Buades, Antoni and Mateu-Lucena, Maite and Jaume, Rafel",
    title = "{Multimode frequency-domain model for the gravitational wave signal from nonprecessing black-hole binaries}",
    eprint = "2001.10914",
    archivePrefix = "arXiv",
    primaryClass = "gr-qc",
    doi = "10.1103/PhysRevD.102.064002",
    journal = "Phys. Rev. D",
    volume = "102",
    number = "6",
    pages = "064002",
    year = "2020"
}

@article{BHPTNRSur1dq1e4,
    author = "Islam, Tousif and Field, Scott E. and Hughes, Scott A. and Khanna, Gaurav and Varma, Vijay and Giesler, Matthew and Scheel, Mark A. and Kidder, Lawrence E. and Pfeiffer, Harald P.",
    title = "{Surrogate model for gravitational wave signals from nonspinning, comparable-to large-mass-ratio black hole binaries built on black hole perturbation theory waveforms calibrated to numerical relativity}",
    eprint = "2204.01972",
    archivePrefix = "arXiv",
    primaryClass = "gr-qc",
    doi = "10.1103/PhysRevD.106.104025",
    journal = "Phys. Rev. D",
    volume = "106",
    number = "10",
    pages = "104025",
    year = "2022"
}

@article{wave_expand,
    author = {Newman, E. T. and Penrose, R.},
    title = "{Note on the Bondi-Metzner-Sachs Group}",
    journal = {Journal of Mathematical Physics},
    volume = {7},
    number = {5},
    pages = {863-870},
    year = {1966},
    month = {05},
    issn = {0022-2488},
    doi = {10.1063/1.1931221},
    url = {https://doi.org/10.1063/1.1931221},
    eprint = {https://pubs.aip.org/aip/jmp/article-pdf/7/5/863/19201010/863\_1\_online.pdf},
}

@article{shoemaker_higher_modes,
  title = {Impact of higher-order modes on the detection of binary black hole coalescences},
  author = {Pekowsky, Larne and Healy, James and Shoemaker, Deirdre and Laguna, Pablo},
  journal = {Phys. Rev. D},
  volume = {87},
  issue = {8},
  pages = {084008},
  numpages = {13},
  year = {2013},
  month = {Apr},
  publisher = {American Physical Society},
  doi = {10.1103/PhysRevD.87.084008},
  url = {https://link.aps.org/doi/10.1103/PhysRevD.87.084008}
}

@article{CalderonBustillo:2015lrt,
    author = {Calder\'on Bustillo, Juan and Husa, Sascha and Sintes, Alicia M. and P\"urrer, Michael},
    title = "{Impact of gravitational radiation higher order modes on single aligned-spin gravitational wave searches for binary black holes}",
    eprint = "1511.02060",
    archivePrefix = "arXiv",
    primaryClass = "gr-qc",
    reportNumber = "LIGO-P1500184",
    doi = "10.1103/PhysRevD.93.084019",
    journal = "Phys. Rev. D",
    volume = "93",
    number = "8",
    pages = "084019",
    year = "2016"
}

@article{ff_intro,
    author = "Apostolatos, T. A.",
    title = "{Search templates for gravitational waves from precessing, inspiraling binaries}",
    doi = "10.1103/PhysRevD.52.605",
    journal = "Phys. Rev. D",
    volume = "52",
    pages = "605--620",
    year = "1995"
}

@article{scipy,
    author = "Virtanen, Pauli and others",
    title = "{SciPy 1.0--Fundamental Algorithms for Scientific Computing in Python}",
    eprint = "1907.10121",
    archivePrefix = "arXiv",
    primaryClass = "cs.MS",
    doi = "10.1038/s41592-019-0686-2",
    journal = "Nature Meth.",
    volume = "17",
    pages = "261",
    year = "2020"
}

@article{measuring_gw_modes,
    author = "Mills, Cameron and Fairhurst, Stephen",
    title = "{Measuring gravitational-wave higher-order multipoles}",
    eprint = "2007.04313",
    archivePrefix = "arXiv",
    primaryClass = "gr-qc",
    doi = "10.1103/PhysRevD.103.024042",
    journal = "Phys. Rev. D",
    volume = "103",
    number = "2",
    pages = "024042",
    year = "2021"
}

@article{constrain_incl,
    author = "Usman, Samantha A. and Mills, Joseph C. and Fairhurst, Stephen",
    title = "{Constraining the Inclinations of Binary Mergers from Gravitational-wave Observations}",
    eprint = "1809.10727",
    archivePrefix = "arXiv",
    primaryClass = "gr-qc",
    doi = "10.3847/1538-4357/ab0b3e",
    journal = "Astrophys. J.",
    volume = "877",
    number = "2",
    pages = "82",
    year = "2019"
}

@article{Kidder:1995zr,
    author = "Kidder, Lawrence E.",
    title = "{Coalescing binary systems of compact objects to postNewtonian 5/2 order. 5. Spin effects}",
    eprint = "gr-qc/9506022",
    archivePrefix = "arXiv",
    reportNumber = "NU-GR-11, WUGRAV-94-6A",
    doi = "10.1103/PhysRevD.52.821",
    journal = "Phys. Rev. D",
    volume = "52",
    pages = "821--847",
    year = "1995"
}

@article{Varma:2014jxa,
    author = {Varma, Vijay and Ajith, Parameswaran and Husa, Sascha and Bustillo, Juan Calderon and Hannam, Mark and P\"urrer, Michael},
    title = "{Gravitational-wave observations of binary black holes: Effect of nonquadrupole modes}",
    eprint = "1409.2349",
    archivePrefix = "arXiv",
    primaryClass = "gr-qc",
    reportNumber = "LIGO-P1400095-V3",
    doi = "10.1103/PhysRevD.90.124004",
    journal = "Phys. Rev. D",
    volume = "90",
    number = "12",
    pages = "124004",
    year = "2014"
}

@article{diff_evo_algo,
    author = "Storn, Rainer and Price, Kenneth",
    title = "{Differential Evolution \textendash{} A Simple and Efficient Heuristic for global Optimization over Continuous Spaces}",
    doi = "10.1023/A:1008202821328",
    journal = "J. Global Optim.",
    volume = "11",
    number = "4",
    pages = "341--359",
    year = "1997"
}

@misc{gwtc4_methods,
    author = "Abac, A. G. and others",
    collaboration = "LIGO Scientific, VIRGO, KAGRA",
    title = "{GWTC-4.0: Methods for Identifying and Characterizing Gravitational-wave Transients}",
    eprint = "2508.18081",
    archivePrefix = "arXiv",
    primaryClass = "gr-qc",
    reportNumber = "LIGO-P2400300",
    month = "8",
    year = "2025"
}

@misc{gwtc4_catalog,
    author = "Abac, A. G. and others",
    collaboration = "LIGO Scientific, VIRGO, KAGRA",
    title = "{GWTC-4.0: Updating the Gravitational-Wave Transient Catalog with Observations from the First Part of the Fourth LIGO-Virgo-KAGRA Observing Run}",
    eprint = "2508.18082",
    archivePrefix = "arXiv",
    primaryClass = "gr-qc",
    reportNumber = "LIGO-P2400386",
    month = "8",
    year = "2025"
}

@misc{gwtc4_open_data,
    author = "Abac, A. G. and others",
    collaboration = "LIGO Scientific, VIRGO, KAGRA",
    title = "{Open Data from LIGO, Virgo, and KAGRA through the First Part of the Fourth Observing Run}",
    eprint = "2508.18079",
    archivePrefix = "arXiv",
    primaryClass = "gr-qc",
    reportNumber = "LIGO-P2500167",
    month = "8",
    year = "2025"
}

@misc{gwtc4_rnp,
    author = "Abac, A. G. and others",
    collaboration = "LIGO Scientific, VIRGO, KAGRA",
    title = "{GWTC-4.0: Population Properties of Merging Compact Binaries}",
    eprint = "2508.18083",
    archivePrefix = "arXiv",
    primaryClass = "astro-ph.HE",
    reportNumber = "LIGO-P2400004",
    month = "8",
    year = "2025"
}

@article{Oppenheimer:1939ue,
    author = "Oppenheimer, J. R. and Snyder, H.",
    title = "{On Continued gravitational contraction}",
    doi = "10.1103/PhysRev.56.455",
    journal = "Phys. Rev.",
    volume = "56",
    pages = "455--459",
    year = "1939"
}

@article{Chandrasekhar:1984zz,
    author = "Chandrasekhar, Subrahmanyan",
    title = "{On stars, their evolution and their stability}",
    doi = "10.1103/RevModPhys.56.137",
    journal = "Rev. Mod. Phys.",
    volume = "56",
    pages = "137--147",
    year = "1984"
}

@ARTICLE{Chandrasekhar1931,
       author = {{Chandrasekhar}, S.},
        title = "{The Maximum Mass of Ideal White Dwarfs}",
      journal = {\apj},
         year = 1931,
        month = jul,
       volume = {74},
        pages = {81},
          doi = {10.1086/143324},
       adsurl = {https://ui.adsabs.harvard.edu/abs/1931ApJ....74...81C}
}

@article{Carr:1975qj,
    author = "Carr, Bernard J.",
    title = "{The Primordial black hole mass spectrum}",
    doi = "10.1086/153853",
    journal = "Astrophys. J.",
    volume = "201",
    pages = "1--19",
    year = "1975"
}

@article{Khlopov:1980mg,
    author = "Khlopov, M. Yu. and Polnarev, A. G.",
    title = "{PRIMORDIAL BLACK HOLES AS A COSMOLOGICAL TEST OF GRAND UNIFICATION}",
    doi = "10.1016/0370-2693(80)90624-3",
    journal = "Phys. Lett. B",
    volume = "97",
    pages = "383--387",
    year = "1980"
}

@article{dasgupta_pbhs,
  title = {Low Mass Black Holes from Dark Core Collapse},
  author = {Dasgupta, Basudeb and Laha, Ranjan and Ray, Anupam},
  journal = {Phys. Rev. Lett.},
  volume = {126},
  issue = {14},
  pages = {141105},
  numpages = {8},
  year = {2021},
  month = {Apr},
  publisher = {American Physical Society},
  doi = {10.1103/PhysRevLett.126.141105},
  url = {https://link.aps.org/doi/10.1103/PhysRevLett.126.141105}
}

@article{Baldes:2023rqv,
    author = "Baldes, Iason and Olea-Romacho, Mar{\'\i}a Olalla",
    title = "{Primordial black holes as dark matter: interferometric tests of phase transition origin}",
    eprint = "2307.11639",
    archivePrefix = "arXiv",
    primaryClass = "hep-ph",
    doi = "10.1007/JHEP01(2024)133",
    journal = "JHEP",
    volume = "01",
    pages = "133",
    year = "2024"
}

@misc{Afroz:2025urb,
    author = "Afroz, Samsuzzaman and Mukherjee, Suvodip",
    title = "{Gravitational Wave Burst from Bremsstrahlung in Milky Way Can Discover Sub-Solar Dark Matter in Near Future}",
    eprint = "2507.22126",
    archivePrefix = "arXiv",
    primaryClass = "astro-ph.CO",
    month = "7",
    year = "2025"
}

@article{Clesse:2016vqa,
    author = "Clesse, Sebastien and Garc{\'\i}a-Bellido, Juan",
    title = "{The clustering of massive Primordial Black Holes as Dark Matter: measuring their mass distribution with Advanced LIGO}",
    eprint = "1603.05234",
    archivePrefix = "arXiv",
    primaryClass = "astro-ph.CO",
    reportNumber = "TTK-16-10, IFT-UAM-CSIC-16-027",
    doi = "10.1016/j.dark.2016.10.002",
    journal = "Phys. Dark Univ.",
    volume = "15",
    pages = "142--147",
    year = "2017"
}

@article{Mukherjee:2021itf,
    author = "Mukherjee, Suvodip and Meinema, Matthew S. P. and Silk, Joseph",
    title = "{Prospects of discovering subsolar primordial black holes using the stochastic gravitational wave background from third-generation detectors}",
    eprint = "2107.02181",
    archivePrefix = "arXiv",
    primaryClass = "astro-ph.CO",
    doi = "10.1093/mnras/stab3756",
    journal = "Mon. Not. Roy. Astron. Soc.",
    volume = "510",
    number = "4",
    pages = "6218--6224",
    year = "2022"
}

@article{Afroz:2024fzp,
    author = "Afroz, Samsuzzaman and Mukherjee, Suvodip",
    title = "{Phase space of binary black holes from gravitational wave observations to unveil its formation history}",
    eprint = "2411.07304",
    archivePrefix = "arXiv",
    primaryClass = "astro-ph.HE",
    doi = "10.1103/7zc2-g9vq",
    journal = "Phys. Rev. D",
    volume = "112",
    number = "2",
    pages = "023531",
    year = "2025"
}

@article{BHPTNRSur2dq1e3,
    author = "Rink, Katie and Bachhar, Ritesh and Islam, Tousif and Rifat, Nur E. M. and Gonzalez-Quesada, Kevin and Field, Scott E. and Khanna, Gaurav and Hughes, Scott A. and Varma, Vijay",
    title = "{Gravitational wave surrogate model for spinning, intermediate mass ratio binaries based on perturbation theory and numerical relativity}",
    eprint = "2407.18319",
    archivePrefix = "arXiv",
    primaryClass = "gr-qc",
    doi = "10.1103/PhysRevD.110.124069",
    journal = "Phys. Rev. D",
    volume = "110",
    number = "12",
    pages = "124069",
    year = "2024"
}

@article{ssm1,
    author = "Abbott, R. and others",
    collaboration = "LVK",
    title = "{Search for subsolar-mass black hole binaries in the second part of Advanced LIGO{\textquoteright}s and Advanced Virgo{\textquoteright}s third observing run}",
    eprint = "2212.01477",
    archivePrefix = "arXiv",
    primaryClass = "astro-ph.HE",
    doi = "10.1093/mnras/stad588",
    journal = "Mon. Not. Roy. Astron. Soc.",
    volume = "524",
    number = "4",
    pages = "5984--5992",
    year = "2023",
    note = "[Erratum: Mon.Not.Roy.Astron.Soc. 526, 6234 (2023)]"
}

@article{ssm3,
    author = "Nitz, Alexander H. and Wang, Yi-Fan",
    title = "{Search for Gravitational Waves from the Coalescence of Subsolar-Mass Binaries in the First Half of Advanced LIGO and Virgo{\textquoteright}s Third Observing Run}",
    eprint = "2106.08979",
    archivePrefix = "arXiv",
    primaryClass = "astro-ph.HE",
    doi = "10.1103/PhysRevLett.127.151101",
    journal = "Phys. Rev. Lett.",
    volume = "127",
    number = "15",
    pages = "151101",
    year = "2021"
}

@article{ssm5,
    author = "Barsanti, Susanna and De Luca, Valerio and Maselli, Andrea and Pani, Paolo",
    title = "{Detecting Subsolar-Mass Primordial Black Holes in Extreme Mass-Ratio Inspirals with LISA and Einstein Telescope}",
    eprint = "2109.02170",
    archivePrefix = "arXiv",
    primaryClass = "gr-qc",
    doi = "10.1103/PhysRevLett.128.111104",
    journal = "Phys. Rev. Lett.",
    volume = "128",
    number = "11",
    pages = "111104",
    year = "2022"
}

@ARTICLE{Carr1974,
       author = {{Carr}, B.~J. and {Hawking}, S.~W.},
        title = "{Black holes in the early Universe}",
      journal = {mnras},
         year = 1974,
        month = aug,
       volume = {168},
        pages = {399-416},
          doi = {10.1093/mnras/168.2.399},
       adsurl = {https://ui.adsabs.harvard.edu/abs/1974MNRAS.168..399C}
}

@article{bird2016,
  title = {Did LIGO Detect Dark Matter?},
  author = {Bird, Simeon and Cholis, Ilias and Mu\~noz, Julian B. and Ali-Ha\"{\i}moud, Yacine and Kamionkowski, Marc and Kovetz, Ely D. and Raccanelli, Alvise and Riess, Adam G.},
  journal = {Phys. Rev. Lett.},
  volume = {116},
  issue = {20},
  pages = {201301},
  numpages = {6},
  year = {2016},
  month = {May},
  publisher = {American Physical Society},
  doi = {10.1103/PhysRevLett.116.201301},
  url = {https://link.aps.org/doi/10.1103/PhysRevLett.116.201301}
}

@article{sasaki2016,
  title = {Primordial Black Hole Scenario for the Gravitational-Wave Event GW150914},
  author = {Sasaki, Misao and Suyama, Teruaki and Tanaka, Takahiro and Yokoyama, Shuichiro},
  journal = {Phys. Rev. Lett.},
  volume = {117},
  issue = {6},
  pages = {061101},
  numpages = {5},
  year = {2016},
  month = {Aug},
  publisher = {American Physical Society},
  doi = {10.1103/PhysRevLett.117.061101},
  url = {https://link.aps.org/doi/10.1103/PhysRevLett.117.061101}
}

@article{Franciolini2022,
  title = {Hunt for light primordial black hole dark matter with ultrahigh-frequency gravitational waves},
  author = {Franciolini, Gabriele and Maharana, Anshuman and Muia, Francesco},
  journal = {Phys. Rev. D},
  volume = {106},
  issue = {10},
  pages = {103520},
  numpages = {30},
  year = {2022},
  month = {Nov},
  publisher = {American Physical Society},
  doi = {10.1103/PhysRevD.106.103520},
  url = {https://link.aps.org/doi/10.1103/PhysRevD.106.103520}
}

@ARTICLE{Jedamzik2020,
       author = {{Jedamzik}, Karsten},
        title = "{Primordial black hole dark matter and the LIGO/Virgo observations}",
      journal = {jcap},
         year = 2020,
        month = sep,
       volume = {2020},
       number = {9},
          eid = {022},
        pages = {022},
          doi = {10.1088/1475-7516/2020/09/022},
archivePrefix = {arXiv},
       eprint = {2006.11172},
 primaryClass = {astro-ph.CO},
       adsurl = {https://ui.adsabs.harvard.edu/abs/2020JCAP...09..022J}
}

@ARTICLE{byrnes2018,
       author = {{Byrnes}, Christian T. and {Hindmarsh}, Mark and {Young}, Sam and {Hawkins}, Michael R.~S.},
        title = "{Primordial black holes with an accurate QCD equation of state}",
      journal = {jcap},
         year = 2018,
        month = aug,
       volume = {2018},
       number = {8},
          eid = {041},
        pages = {041},
          doi = {10.1088/1475-7516/2018/08/041},
archivePrefix = {arXiv},
       eprint = {1801.06138},
 primaryClass = {astro-ph.CO},
       adsurl = {https://ui.adsabs.harvard.edu/abs/2018JCAP...08..041B}
}

@ARTICLE{Prunier2023SSM200308,
       author = {{Prunier}, Marine and {Morr{\'a}s}, Gonzalo and {Siles}, Jos{\'e} Francisco Nu{\~n}o and {Clesse}, Sebastien and {Garc{\'\i}a-Bellido}, Juan and {Morales}, Ester Ruiz},
        title = "{Analysis of the subsolar-mass black hole candidate SSM200308 from the second part of the third observing run of Advanced LIGO-Virgo}",
      journal = {Physics of the Dark Universe},
         year = 2024,
        month = dec,
       volume = {46},
          eid = {101582},
        pages = {101582},
          doi = {10.1016/j.dark.2024.101582},
archivePrefix = {arXiv},
       eprint = {2311.16085},
 primaryClass = {gr-qc},
       adsurl = {https://ui.adsabs.harvard.edu/abs/2024PDU....4601582P}
}

@misc{Amaro2017,
    author = "Amaro-Seoane, Pau and others",
    collaboration = "LISA",
    title = "{Laser Interferometer Space Antenna}",
    eprint = "1702.00786",
    archivePrefix = "arXiv",
    primaryClass = "astro-ph.IM",
    month = "2",
    year = "2017"
}

@article{Amaro2018,
    author = "Amaro-Seoane, Pau",
    title = "{Detecting Intermediate-Mass Ratio Inspirals From The Ground And Space}",
    eprint = "1807.03824",
    archivePrefix = "arXiv",
    primaryClass = "astro-ph.HE",
    doi = "10.1103/PhysRevD.98.063018",
    journal = "Phys. Rev. D",
    volume = "98",
    number = "6",
    pages = "063018",
    year = "2018"
}

@article{islam2021,
    author = "Islam, Tousif and Field, Scott E. and Haster, Carl-Johan and Smith, Rory",
    title = "{High precision source characterization of intermediate mass-ratio black hole coalescences with gravitational waves: The importance of higher order multipoles}",
    eprint = "2105.04422",
    archivePrefix = "arXiv",
    primaryClass = "gr-qc",
    reportNumber = "This is LIGO Document Number DCC-P2100151",
    doi = "10.1103/PhysRevD.104.084068",
    journal = "Phys. Rev. D",
    volume = "104",
    number = "8",
    pages = "084068",
    year = "2021"
}

@ARTICLE{Torres2025,
       author = {{Torres-Orjuela}, Alejandro and {V{\'a}zquez-Aceves}, Ver{\'o}nica and {Wang}, Tian-Xiao},
        title = "{Detection of Intermediate-mass Ratio Inspirals in Globular Clusters: Revealing the Brownian Motion with Gravitational Waves}",
      journal = {\apj},
         year = 2025,
        month = jun,
       volume = {986},
       number = {2},
          eid = {155},
        pages = {155},
          doi = {10.3847/1538-4357/add5f4},
archivePrefix = {arXiv},
       eprint = {2501.13466},
 primaryClass = {astro-ph.HE},
       adsurl = {https://ui.adsabs.harvard.edu/abs/2025ApJ...986..155T}
}

@misc{Cheung2025IMRI,
    author = "Cheung, Mark Ho-Yeuk and Wadekar, Digvijay and Mehta, Ajit Kumar and Islam, Tousif and Roulet, Javier and Berti, Emanuele and Venumadhav, Tejaswi and Zackay, Barak and Zaldarriaga, Matias",
    title = "{Searching for intermediate mass ratio binary black hole mergers in the third observing run of LIGO-Virgo-KAGRA}",
    eprint = "2507.01083",
    archivePrefix = "arXiv",
    primaryClass = "gr-qc",
    month = "7",
    year = "2025"
}

@article{Hanna2024SSMBank,
    author = "Hanna, Chad and others",
    title = "{Template bank for subsolar mass compact binary mergers in the fourth observing run of Advanced LIGO, Advanced Virgo, and KAGRA}",
    eprint = "2412.10951",
    archivePrefix = "arXiv",
    primaryClass = "gr-qc",
    doi = "10.1103/c97v-bmj8",
    journal = "Phys. Rev. D",
    volume = "112",
    number = "4",
    pages = "044013",
    year = "2025"
}

@article{Estelles2022,
    author = "Estell{\'e}s, H{\'e}ctor and Colleoni, Marta and Garc{\'\i}a-Quir{\'o}s, Cecilio and Husa, Sascha and Keitel, David and Mateu-Lucena, Maite and Planas, Maria de Lluc and Ramos-Buades, Antoni",
    title = "{New twists in compact binary waveform modeling: A fast time-domain model for precession}",
    eprint = "2105.05872",
    archivePrefix = "arXiv",
    primaryClass = "gr-qc",
    reportNumber = "LIGO-P2100136",
    doi = "10.1103/PhysRevD.105.084040",
    journal = "Phys. Rev. D",
    volume = "105",
    number = "8",
    pages = "084040",
    year = "2022"
}

@article{Crescimbeni2024Tidal,
    author = "Crescimbeni, Francesco and Franciolini, Gabriele and Pani, Paolo and Riotto, Antonio",
    title = "{Can we identify primordial black holes? Tidal tests for subsolar-mass gravitational-wave observations}",
    eprint = "2402.18656",
    archivePrefix = "arXiv",
    primaryClass = "astro-ph.HE",
    reportNumber = "CERN-TH-2024-026",
    doi = "10.1103/PhysRevD.109.124063",
    journal = "Phys. Rev. D",
    volume = "109",
    number = "12",
    pages = "124063",
    year = "2024"
}

@ARTICLE{Maggiore2020,
       author = {{Maggiore}, Michele et al.},
        title = "{Science case for the Einstein telescope}",
      journal = {jcap},
         year = 2020,
        month = mar,
       volume = {2020},
       number = {3},
          eid = {050},
        pages = {050},
          doi = {10.1088/1475-7516/2020/03/050},
archivePrefix = {arXiv},
       eprint = {1912.02622},
 primaryClass = {astro-ph.CO},
       adsurl = {https://ui.adsabs.harvard.edu/abs/2020JCAP...03..050M}
}

@ARTICLE{Yuan2024PBH,
       author = {{Yuan}, Chen and {Huang}, Qing-Guo},
        title = "{Primordial black hole interpretation in subsolar mass gravitational wave candidate SSM200308}",
      journal = {jcap},
         year = 2024,
        month = sep,
       volume = {2024},
       number = {9},
          eid = {051},
        pages = {051},
          doi = {10.1088/1475-7516/2024/09/051},
archivePrefix = {arXiv},
       eprint = {2404.03328},
 primaryClass = {astro-ph.CO},
       adsurl = {https://ui.adsabs.harvard.edu/abs/2024JCAP...09..051Y}
}

@article{Berti2015,
    author = "Berti, Emanuele and others",
    title = "{Testing General Relativity with Present and Future Astrophysical Observations}",
    eprint = "1501.07274",
    archivePrefix = "arXiv",
    primaryClass = "gr-qc",
    doi = "10.1088/0264-9381/32/24/243001",
    journal = "Class. Quant. Grav.",
    volume = "32",
    pages = "243001",
    year = "2015"
}

@article{Yunes2016,
    author = "Yunes, Nicolas and Yagi, Kent and Pretorius, Frans",
    title = "{Theoretical Physics Implications of the Binary Black-Hole Mergers GW150914 and GW151226}",
    eprint = "1603.08955",
    archivePrefix = "arXiv",
    primaryClass = "gr-qc",
    doi = "10.1103/PhysRevD.94.084002",
    journal = "Phys. Rev. D",
    volume = "94",
    number = "8",
    pages = "084002",
    year = "2016"
}

@article{Shadykul2023,
    title={Intermediate mass ratio inspirals in dark matter halos},
    author={Shadykul, Darkhan and Chakrabarty, Hrishikesh and Malafarina, Daniele},
    journal={Phys. Rev. D},
    volume={111},
    number={10},
    pages={104003},
    year={2025},
    publisher={American Physical Society},
    doi={10.1103/PhysRevD.111.104003}
}

@misc{PyCBC,
  doi = {10.5281/ZENODO.596388},
  url = {https://zenodo.org/doi/10.5281/zenodo.596388},
  author = "Alex Nitz and others",
  title = {gwastro/pycbc: v2.3.3 release of PyCBC},
  publisher = {Zenodo},
  year = {2024},
  copyright = {Creative Commons Attribution 4.0 International}
}

@Article{         numpy,
 title         = {Array programming with {NumPy}},
 author        = {Charles R. Harris and K. Jarrod Millman and St{\'{e}}fan J.
                 van der Walt and Ralf Gommers and Pauli Virtanen and David
                 Cournapeau and Eric Wieser and Julian Taylor and Sebastian
                 Berg and Nathaniel J. Smith and Robert Kern and Matti Picus
                 and Stephan Hoyer and Marten H. van Kerkwijk and Matthew
                 Brett and Allan Haldane and Jaime Fern{\'{a}}ndez del
                 R{\'{i}}o and Mark Wiebe and Pearu Peterson and Pierre
                 G{\'{e}}rard-Marchant and Kevin Sheppard and Tyler Reddy and
                 Warren Weckesser and Hameer Abbasi and Christoph Gohlke and
                 Travis E. Oliphant},
 year          = {2020},
 month         = sep,
 journal       = {Nature},
 volume        = {585},
 number        = {7825},
 pages         = {357--362},
 doi           = {10.1038/s41586-020-2649-2},
 publisher     = {Springer Science and Business Media {LLC}},
 url           = {https://doi.org/10.1038/s41586-020-2649-2}
}

@article{bilby_paper,
    author = "Ashton, Gregory and others",
    title = "{BILBY: A user-friendly Bayesian inference library for gravitational-wave astronomy}",
    eprint = "1811.02042",
    archivePrefix = "arXiv",
    primaryClass = "astro-ph.IM",
    doi = "10.3847/1538-4365/ab06fc",
    journal = "Astrophys. J. Suppl.",
    volume = "241",
    number = "2",
    pages = "27",
    year = "2019"
}

@Article{matplotlib,
  Author    = {Hunter, J. D.},
  Title     = {Matplotlib: A 2D graphics environment},
  Journal   = {Computing in Science \& Engineering},
  Volume    = {9},
  Number    = {3},
  Pages     = {90--95},
  publisher = {IEEE COMPUTER SOC},
  doi       = {10.1109/MCSE.2007.55},
  year      = 2007
}

@article{Pratten2020,
    title={Setting the cornerstone for the IMRPhenomX family of models for gravitational waves from compact binaries: The dominant harmonic for non-precessing quasi-circular black holes},
    author={Pratten, Geraint and Husa, Sascha and García-Quirós, Cecilio and Colleoni, Marta and Ramos-Buades, Antoni and Estellés, Héctor and Jaume, Rafel},
    journal={Phys. Rev. D},
    volume={102},
    number={6},
    pages={064001},
    year={2020},
    month={Sep},
    publisher={American Physical Society},
    doi={10.1103/PhysRevD.102.064001},
    eprint={2001.11412},
    archivePrefix={arXiv},
    primaryClass={gr-qc}
}

@article{GarciaQuiros2020,
    title={IMRPhenomXHM: A multi-mode frequency-domain model for the gravitational wave signal from non-precessing black-hole binaries},
    author={García-Quirós, Cecilio and Colleoni, Marta and Husa, Sascha and Estellés, Héctor and Pratten, Geraint and Ramos-Buades, Antoni and Mateu-Lucena, Maite and Jaume, Rafel},
    journal={Phys. Rev. D},
    volume={102},
    number={6},
    pages={064002},
    year={2020},
    month={Sep},
    publisher={American Physical Society},
    doi={10.1103/PhysRevD.102.064002},
    eprint={2001.10914},
    archivePrefix={arXiv},
    primaryClass={gr-qc}
}

@article{Colleoni2023,
    title={Accurate and efficient waveform model for precessing binary black holes},
    author={Colleoni, Marta and others},
    journal={Phys. Rev. D},
    volume={108},
    number={6},
    pages={064059},
    year={2023},
    month={Sep},
    publisher={American Physical Society},
    doi={10.1103/PhysRevD.108.064059},
    eprint={2303.18039},
    archivePrefix={arXiv},
    primaryClass={gr-qc}
}

@article{Islam2022,
    title={Surrogate model for gravitational wave signals from nonspinning, comparable-to large-mass-ratio black hole binaries built on black hole perturbation theory waveforms calibrated to numerical relativity},
    author={Islam, Tousif and Field, Scott E. and Hughes, Scott A. and Khanna, Gaurav and Varma, Vijay and Giesler, Matthew and Scheel, Mark A. and Kidder, Lawrence E. and Pfeiffer, Harald P.},
    journal={Phys. Rev. D},
    volume={106},
    number={10},
    pages={104025},
    year={2022},
    month={Nov},
    publisher={American Physical Society},
    doi={10.1103/PhysRevD.106.104025},
    eprint={2204.01972},
    archivePrefix={arXiv},
    primaryClass={gr-qc}
}

@article{Boyle2019,
    title={The SXS Collaboration catalog of binary black hole simulations},
    author={Boyle, Michael and others},
    collaboration={SXS Collaboration},
    journal={Class. Quantum Grav.},
    volume={36},
    number={19},
    pages={195006},
    year={2019},
    month={Sep},
    publisher={IOP Publishing},
    doi={10.1088/1361-6382/ab34e2},
    eprint={1904.04831},
    archivePrefix={arXiv},
    primaryClass={gr-qc}
}

@article{Healy2017,
    title={Numerical relativity simulations of spinning black hole binaries with precession and higher modes: Application to GW151226},
    author={Healy, James and Lousto, Carlos O. and Zlochower, Yosef},
    journal={Phys. Rev. D},
    volume={96},
    number={2},
    pages={024031},
    year={2017},
    month={Jul},
    publisher={American Physical Society},
    doi={10.1103/PhysRevD.96.024031},
    eprint={1705.07034},
    archivePrefix={arXiv},
    primaryClass={gr-qc}
}

@article{Teukolsky1973,
    title={Perturbations of a Rotating Black Hole. I. Fundamental Equations for Gravitational, Electromagnetic, and Neutrino-Field Perturbations},
    author={Teukolsky, Saul A.},
    journal={Astrophys. J.},
    volume={185},
    pages={635--647},
    year={1973},
    month={Oct},
    doi={10.1086/152444}
}

@article{Hughes2000,
    title={Evolution of circular, nonequatorial orbits of Kerr black holes due to gravitational-wave emission},
    author={Hughes, Scott A.},
    journal={Phys. Rev. D},
    volume={61},
    pages={084004},
    year={2000},
    month={Mar},
    publisher={American Physical Society},
    doi={10.1103/PhysRevD.61.084004},
    eprint={gr-qc/9910091},
    archivePrefix={arXiv},
    primaryClass={gr-qc}
}

@article{Islam2023cal,
    title={On the approximate relation between black-hole perturbation theory and numerical relativity},
    author={Islam, Tousif and Khanna, Gaurav},
    journal={Phys. Rev. D},
    volume={108},
    number={12},
    pages={124046},
    year={2023},
    month={Dec},
    publisher={American Physical Society},
    doi={10.1103/PhysRevD.108.124046},
    eprint={2307.03155},
    archivePrefix={arXiv},
    primaryClass={gr-qc}
}

@article{Islam2023finite,
    title={Interplay between numerical relativity and black hole perturbation theory: Finite size effects},
    author={Islam, Tousif and Khanna, Gaurav},
    journal={Phys. Rev. D},
    volume={108},
    number={4},
    pages={044013},
    year={2023},
    month={Aug},
    publisher={American Physical Society},
    doi={10.1103/PhysRevD.108.044013},
    eprint={2306.08767},
    archivePrefix={arXiv},
    primaryClass={gr-qc}
}

@article{Field2014,
    title={Fast prediction and evaluation of gravitational waveforms using surrogate models},
    author={Field, Scott E. and Galley, Chad R. and Herrmann, Frank and Hesthaven, Jan S. and Ochsner, Evan and Tiglio, Manuel},
    journal={Phys. Rev. X},
    volume={4},
    number={3},
    pages={031006},
    year={2014},
    month={Jul},
    publisher={American Physical Society},
    doi={10.1103/PhysRevX.4.031006},
    eprint={1308.3565},
    archivePrefix={arXiv},
    primaryClass={gr-qc}
}

@misc{BHPToolkit2024,
    author={{Black Hole Perturbation Toolkit Collaboration}},
    title={Black Hole Perturbation Toolkit},
    howpublished={\url{https://bhptoolkit.org}},
    year={2024},
    note={Accessed: 2024-10-16}
}

@article{Estelles2021,
    title={Waveform accuracy requirements for the Einstein Telescope},
    author={Estellés, Héctor and others},
    journal={Class. Quantum Grav.},
    volume={38},
    number={15},
    pages={155010},
    year={2021},
    month={Jul},
    publisher={IOP Publishing},
    doi={10.1088/1361-6382/ac10e1},
    eprint={2104.01565},
    archivePrefix={arXiv},
    primaryClass={gr-qc}
}

\end{document}